\pdfoutput=1

\documentclass[10pt,pdftex,a4paper]{article}

\usepackage{etex}
\reserveinserts{18}

\usepackage{a4wide}
\usepackage{setspace,xspace,makeidx,fancyvrb,relsize}
\usepackage{alltt}

\usepackage{amsmath,amsthm,amssymb}
\usepackage{parallel,amsfonts,accents}

\usepackage{amsxtra,amstext,latexsym,dsfont} 

\normalfont
\usepackage{bm}

\usepackage[comma]{natbib}

\usepackage[symbol]{footmisc}
\usepackage{perpage}
\MakePerPage[2]{footnote}

\usepackage[pdftex]{graphicx}
\usepackage[dvipsnames]{xcolor}

\usepackage{pgf,tikz}
\usetikzlibrary{snakes}
\usetikzlibrary{backgrounds}
\usetikzlibrary{arrows}
\usetikzlibrary{patterns}
\usetikzlibrary{fit}
\usetikzlibrary{calc}

\usepackage{lscape} 
\usepackage{shorttoc}

\usepackage{array,longtable,booktabs,float,rotating}
\usepackage{dcolumn,ctable}

\usepackage[labelfont={bf},ruled,labelsep=period,small]{caption}

\usepackage[flushleft]{threeparttable}

\usepackage[linkcolor=black,colorlinks=true,citecolor=black,
bookmarks=true,bookmarksopen=true,bookmarksnumbered=true,bookmarksopenlevel=0,%
hyperindex=true,pdfpagemode=UseOutlines,linktocpage=true,pdfpagelayout=TwoColumnLeft,
pdftitle=Biostatistics,pdfstartview=FitH,pdffitwindow=true,urlcolor=blue]{hyperref}

\usepackage{ragged2e}

\usepackage{multirow,multicol}

\usepackage{fancyhdr,extramarks}

 \makeatletter
 \newcommand\sub{\@startsection%
     {subsubsection}{5}{0mm}{-1\baselineskip}{.01\baselineskip}%
     {\normalfont\itshape}}
 \renewcommand\subsubsection{\@startsection%
     {subsubsection}{3}{0mm}{-1\baselineskip}{.01\baselineskip}%
     {\normalfont\itshape}}
 \makeatother

\makeatletter
        \newcommand\Appendix[2][?]{%
            \refstepcounter{section}%
            \addcontentsline{toc}{appendix}%
                {\protect\numberline{\appendixname~\thesection}#1}%
            {\raggedleft\bfseries \appendixname\
                \thesection\par \centering#2\par}%
                \sectionmark{#1}%
                \@afterheading
                \addvspace{\baselineskip}}
        \newcommand\sAppendix[1]{%
            \raggedleft\bfseries\appendixname\par
            \@afterheading\addvspace{\baselineskip}}
    \makeatother

\normalsize \makeindex

\newcolumntype{A}{>{\centering}p{100pt}}
\newlength\savedwidth

\def\coldot{.}%
{\catcode`\.=\active%
    \gdef.{$\egroup\setbox2=\hbox to \dimen0 \bgroup$\coldot}}
\def\rightdots#1{%
    \setbox0=\hbox{$1$}\dimen0=#1\wd0%
    \setbox0=\hbox{$\coldot$}\advance\dimen0 \wd0%
    \setbox2=\hbox to \dimen0 {}%
    \setbox0=\hbox\bgroup\mathcode`\.="8000 $}
\def\endrightdots{$\hfil\egroup\box0\box2}

\newcolumntype{d}[1]{D{.}{.}{#1}}
\newcolumntype{A}{>{\centering}p{100pt}}
\newcolumntype{.}{D{.}{.}{-1}}
\newcolumntype{P}[2]{>{#1\raggedright\arraybackslash}p{#2}}

\DeclareFontFamily{U}{euc}{}
\DeclareFontShape{U}{euc}{m}{n}{<-6>eurm5<6-8>eurm7<8->eurm10}{}%

\theoremstyle{plain}      
\theoremstyle{plain}      
\theoremstyle{plain}      
\theoremstyle{definition} 
\theoremstyle{definition} 
\theoremstyle{definition} \newtheorem{exa}{Example}
\theoremstyle{plain} 
\theoremstyle{definition} 
\theoremstyle{plain} \newtheorem{pro}{Proposition}
\theoremstyle{definition} 

\newcounter{nctr}
\newenvironment{3table}{\begin{threeparttable}}{\end{threeparttable}}

\VerbatimFootnotes

\usepackage{url}
\newcommand\tb{\textbf}
\newcommand\ti{\textit}

\newcommand\mcol{\multicolumn}

\newcommand\ds{\mathds}
\newcommand\bb{\mathbb}

\newcommand\te{\text}
\newcommand\ma[1]{\te{\bf{#1}}}
\newcommand\ca{\mathcal}

\newcommand\op{\operatorname}

\newcommand\as{^\ast}

\newcommand\E{\bb{E}}

\newcommand\iid{\op{iid}}

\newcommand\lb{\lbrace}
\newcommand\lt{\left}

\newcommand\pr{^\prime}
\newcommand\ppr{^{\prime\prime}}

\newcommand\qq{\qquad}

\newcommand\rb{\rbrace}
\newcommand\rt{\right}

\newcommand\stack{\stackrel} 

\newcommand\tth{^\text{th}}
\newcommand\var{\operatorname{\bb{V}ar}}

\newcommand\R{\ds{R}}  

\newcommand\Bb{\ma{b}} 
\newcommand\bk{\ma{k}} %
\newcommand\by{\ma{y}}

\newcommand\bA{\ma{A}} 
\newcommand\bI{\ma{I}} 
\newcommand\bP{\ma{P}} 
\newcommand\bR{\ma{R}} 
\newcommand\bW{\ma{W}} 
\newcommand\bX{\ma{X}}

\newcommand\bZ{\ma{Z}}

\newcommand\bone{\bm{1}} 

\newcommand\cE{\ca{E}} 
\newcommand\cI{\ca{I}} 
\newcommand\cO{\ca{O}} %
\newcommand\cP{\ca{P}} 
\newcommand\cV{\ca{V}} 
\newcommand\cW{\ca{W}} %
\newcommand\ga{\gamma}

\newcommand\ep{\epsilon}

\newcommand\sig{\sigma}

\newcommand\bbe{\bm\beta}

\newcommand\bep{\bm\epsilon}

\usepackage{fancyhdr}
\begin{document}
\sloppy

\begin{center}
Running Head: \uppercase{Weighted Network Analysis} 
\end{center}
\vspace{3cm}

\begin{center}
\Large{\tb{Weighted Network Analysis:\\ Separating Cost from Topology using Cost-integration.}}
\\
\vspace{2.5cm} \normalsize 
               Cedric E. Ginestet${^{ab}}$, \\
               Thomas E. Nichols${^{c}}$, 
               Ed T. Bullmore${^{d}}$ 
               and Andrew Simmons${^{ab}}$
\end{center}
\begin{center}
  \vspace{1cm} 
  \rm ${^a}$ King's College London, Institute of Psychiatry,
  Department of Neuroimaging \\
  \rm $^b$ National Institute of Health Research (NIHR) Biomedical
  Research Centre for Mental Health at South London and King's College
  London Institute of Psychiatry \\
  \rm $^c$ Department of Statistics, University of Warwick, Coventry.\\
  \rm $^d$ Brain Mapping Unit, Department of Psychiatry, School of
  Clinical Medicine, \\ University of Cambridge. \\
\end{center}
\vspace{8cm}
Correspondence concerning this article should be sent to Cedric
Ginestet at the Centre for Neuroimaging Sciences, NIHR Biomedical Research Centre,
Institute of Psychiatry, Box P089, King's College London, 
De Crespigny Park, London, SE5 8AF, UK. Email may be sent to 
\textcolor{blue}{\url{cedric.ginestet@kcl.ac.uk}}
\pagebreak






\begin{abstract}
A statistically principled way of conducting weighted network
analysis is still lacking. Comparison of different
populations of weighted networks is hard
because topology is inherently dependent on wiring cost, where cost is
defined as the number of edges in an unweighted graph.
In this paper, we evaluate the benefits and limitations associated
with using cost-integrated topological metrics. 
Our focus is on comparing populations of weighted undirected
graphs using global efficiency. We evaluate different approaches to the comparison of weighted
networks that differ in mean association weight. Our key result
shows that integrating over cost is equivalent to
controlling for any monotonic transformation of the weight set of a
weighted graph. That is, when integrating over cost, we eliminate the
differences in topology that may be due to
a monotonic transformation of the weight set. Our result holds for
any unweighted topological measure. Cost-integration is therefore
helpful in disentangling differences in cost from
differences in topology. By contrast,
we show that the use of the weighted version of a topological metric 
does not constitute a valid approach to this
problem. Indeed, we prove that, under mild conditions, the use of
the weighted version of global efficiency is equivalent to simply comparing
weighted costs. Thus, we recommend the reporting of (i) differences
in weighted costs and (ii) differences in cost-integrated topological measures. 
We demonstrate the application of these techniques in a re-analysis of
an fMRI working memory task. 
We also provide a Monte Carlo method for
approximating cost-integrated topological measures. Finally, we
discuss the limitations of integrating topology over cost, which
may pose problems when some weights are zero, when 
multiplicities exist in the ranks of the weights, and when one expects 
subtle cost-dependent topological differences, which could be masked by
cost-integration. 

\end{abstract}

\section{Introduction}
In the last decade, the biological and physical sciences have
witnessed a proliferation of publications adopting a network approach to a
wide range of questions. This
interest in networks was originally stimulated by the seminal works of
\citet{Watts1998} and \citet{Barabasi1999}, who introduced the
concepts of small-world and scale-free networks, respectively. 
Some of these ideas have been adopted in neuroscience at both a theoretical 
\citep{Sporns2000,Sporns2004} and experimental level \citep{Eguiluz2005}. 
Most of the research in this area has attempted to classify the
topology of brain networks based on anatomical 
or functional data \citep[see, for example,][]{Achard2006,Achard2007,He2007}. 

A question that naturally arises from such applications of
graph theory is whether or not the topological properties of these
brain networks are stable across different populations of subjects or
across different cognitive and behavioral tasks.  A
common hypothesis that neuroscientists may wish to test is whether the
small-world properties of a given brain network are conserved when
comparing patients and healthy controls. \citet{Bassett2008}, for example, have
studied differences in anatomical brain networks between healthy
controls and patients with schizophrenia. Other authors have evaluated whether the 
topological properties of functional networks vary with different
behavioral tasks \citep{Heuvel2009,Fallani2008,
Cecchi2007,Astolfi2009}. The properties of brain network
topology have also been studied at different spatial scales
\citep{Bassett2006} and using different modalities, such as EEG
\citep{Pachou2008,Salvador2008}, and fMRI \citep{Achard2006,Achard2007}. 
There is therefore considerable interest in comparing populations of
networks --which may represent different groups of
subjects, several conditions of an experiment, or the use of different
levels of spatial or temporal resolution. We note that such research
questions are more likely to arise when subject-specific networks can
be directly constructed. This has been done in the context of both
functional and structural MRI \citep{Hagmann2008,Gong2009}. 

The possibility of conducting rigorous statistical 
comparison of several populations of networks, however, has been
hindered by a series of methodological issues, which
have not been hitherto satisfactorily resolved. 
When considering the question of comparing several populations of
networks, two main problems arise. Firstly, we are faced with the
inherent intertwining of connectivity strength (i.e.~ wiring cost)
with network topology. Most topological metrics used to compare
networks are sensitive to differences in these
graphs' number of edges. Drawing comparisons on the sole basis of
topology therefore requires some level of control of cost discrepancies
between these network populations. Secondly, this issue is
compounded by the fundamental division between weighted and unweighted
graphs. The problem of disentangling differences in connectivity
strength from topological
differences therefore needs to be resolved in a distinct manner
depending on whether weighted or unweighted
graphs are being considered. The focus, in this paper, will be on
weighted networks since
these are more likely to be found in the biomedical sciences than their
unweighted counterparts. 

Historically, however, network analyses have concentrated on
unweighted graphs. The application of graph theory to 
biological and artificial networks was originally motivated by the discrete
nature of the problems of interest. Both \citet{Watts1998} and
\citet{Barabasi1999} mainly considered binary relations between sets
of elements, which
readily produced adjacency matrices that could then be used to construct
unweighted graphs. \citet{Watts1998} matched some networks of interest
with their random and regular equivalents. In their case, the matching procedure
ensured that both random and regular networks possessed the same total
number of nodes and edges as
the original graph. Current practice in MRI-based neuroscience and
other biomedical applications, however, tends to produce \ti{weighted} connectivity
networks. This is because MRI data take values on a continuous scale,
which lends itself to the
application of real-valued measures of association, such as the correlation
coefficient or the synchronization likelihood among others. While
different populations of unweighted networks can readily be compared
by matching each network with a random network possessing an identical
number of edges; there is, as yet, no consensus on how to compare populations of
weighted networks in a systematic manner. 

\begin{figure}[t]
  \centering
  \tb{(a)} \\ \vspace{.1cm}
  \includegraphics[width=3cm]{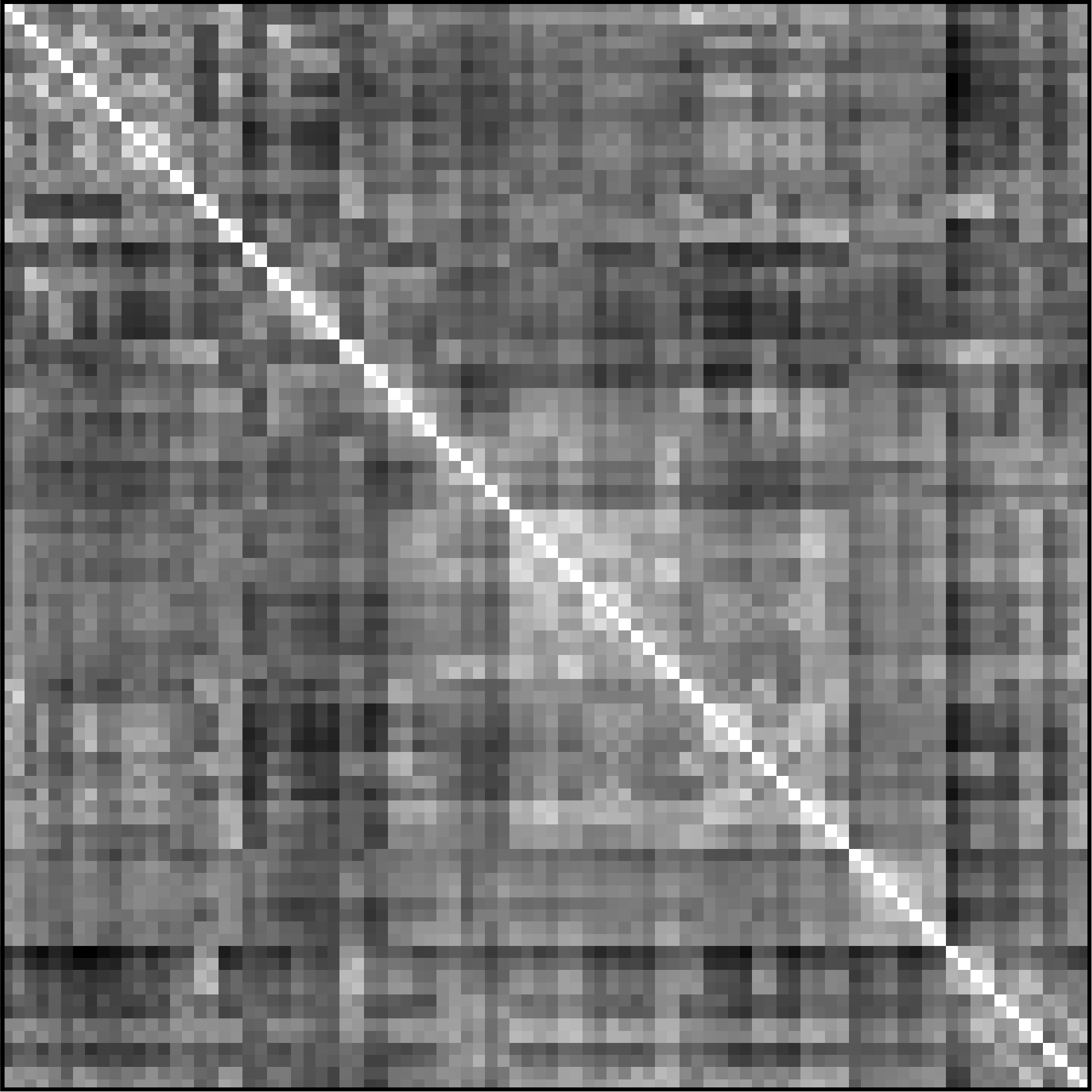}
  \includegraphics[width=3cm]{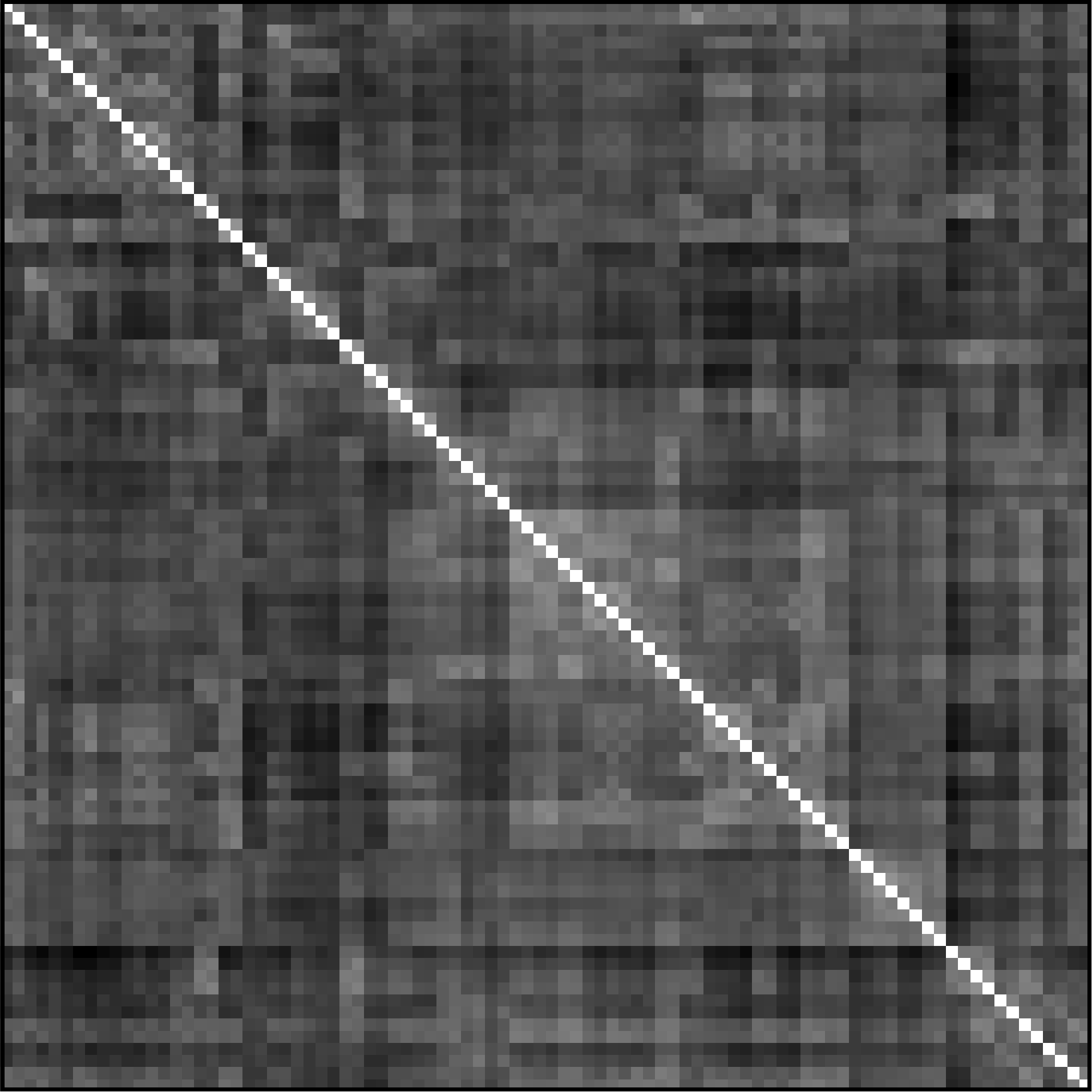} \\
  \tb{(b)} \\ \vspace{.1cm}
  \includegraphics[width=3cm]{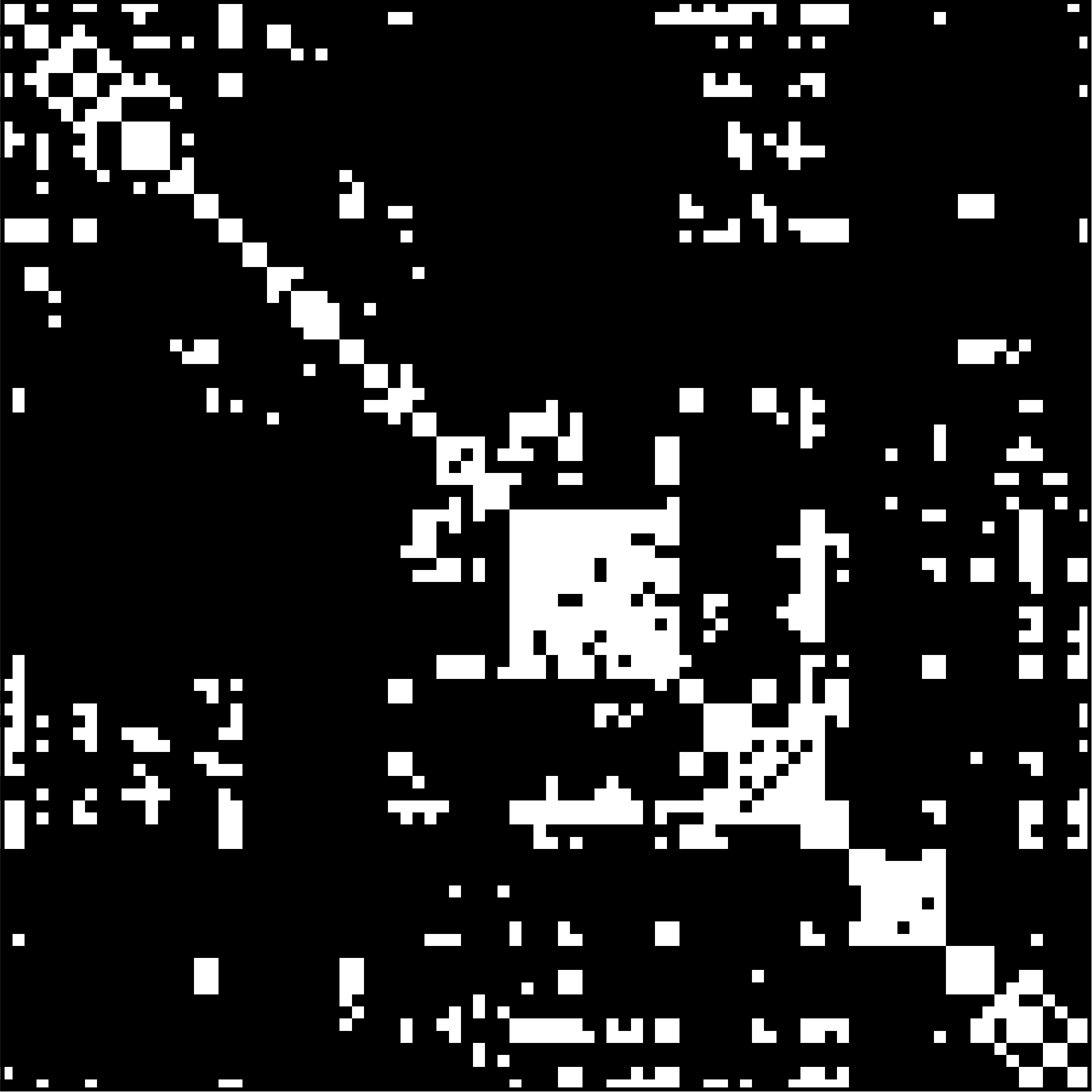}
  \includegraphics[width=3cm]{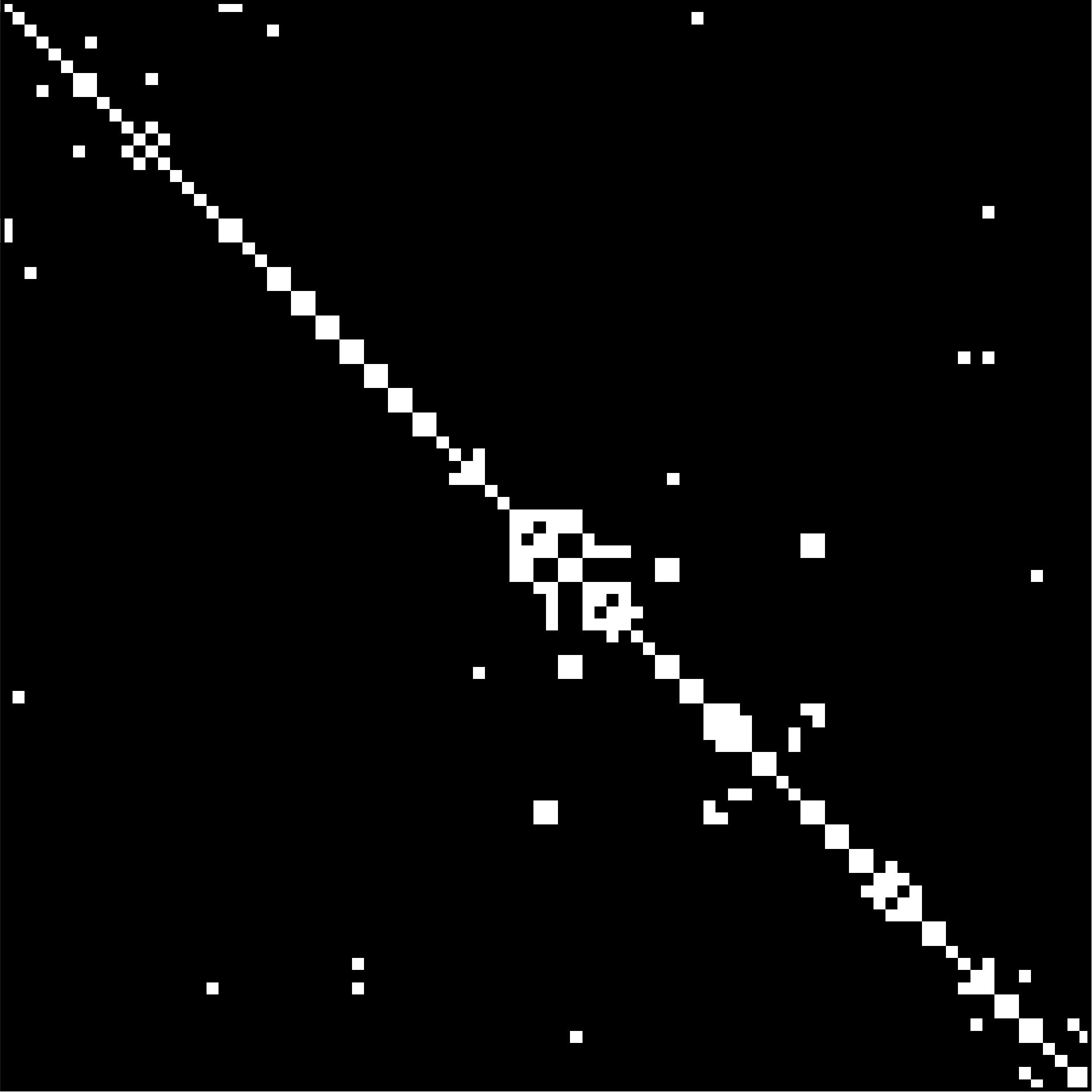} \\
  \vspace{.5cm}
  \caption{How can one disentangle differences in connectivity
    strength from differences in
    topology? In panel (a), two correlation matrices for two weighted
    networks differ in their
    average correlation strengths. In panel (b), the same 
    correlation matrices have been thresholded at the same value,
    producing graphs with different cost levels. In all matrices, black
    indicates null values, and white denotes entries equal to unity.  
    \label{fig:introduction}}
\end{figure}

This problem can be illustrated with a straightforward example. In
panel (a) of Figure \ref{fig:introduction}, a pair of weighted 
networks are represented by their correlation matrices.
We are interested in comparing the topology of the corresponding
weighted graphs. Since these networks differ in their mean correlation
coefficients, a simple thresholding of these matrices will produce
graphs of different levels of cost. Naturally, this is only one of
the different thresholding approaches that could be adopted. 
This non-uniqueness arises since graph topology is
expressed in the language of discrete mathematics, whereas correlation
coefficients are real-valued
functions. That is, one cannot directly adopt concepts originally
developed for unweighted graphs for the analysis of weighted
graphs. 

In this paper, we consider two main approaches to the problem of
weighted network comparison. Firstly, following other authors, we
evaluate the use of weighted topological metrics, which are 
weighted equivalents of graph-theoretical metrics for unweighted
networks \citep{Latora2001,Rubinov2010}. Secondly, we consider 
the utilization of cost-integrated measures of topology, where all the
possible wiring costs of a network are taken into account. When
unweighted, a graph's wiring cost is defined as its number of edges.
Integrating over wiring cost can here be interpreted in statistical
terms, as an analog to the Bayesian integration of nuisance
parameters. Doing so, we are averaging out the `uncertainty' in the choice of
a particular level of cost. This second family of measures has
probably been the most popular to date in the
neuroscientific literature \citep[see][for
instance]{Achard2006,He2009}. However, different authors have chosen
different integration intervals. We therefore explore the consequences
of integrating a topological metric with respect to different subsets of the
cost interval. 

Note that our approach substantially differs from the one adopted by
\citet{Wijk2010}, who proposed several formulas relating cost levels
and topological measures, such as the characteristic path length or
the clustering coefficient. Instead, in this paper, we are concerned
with formally deriving what is the effect of integrating a particular
topological measures over cost levels, in order to assess whether this
is a successful manner of disentangling differences in cost from
differences in topology. In particular, although \citet{Wijk2010}
reviews several ways of controlling for differences in cost, they do
not consider cost-integration, per se. This paper can therefore
be seen as a contribution to the literature on weighted network
analysis, where we formally clarify the utilization of
cost-integration when comparing topological metrics. 

The concept of topology in the context of this paper will be defined
in a quantitative manner. This should be contrasted with the
qualitative definition adopted by previous authors. \citet{Wijk2010},
for instance, assume that networks that represent different
realizations of the same `generative model' should be regarded as
topologically identical. Several realizations from an
Erd\"{o}s-R\'{e}nyi model with fixed edge probability, for example,
share a common generative model and therefore can be said to have an
identical topology. In practice, however, such a generative model is
unknown. Thus, we will refer to this type of classification as a topological
taxon. A taxonony of commonly encountered networks may include
the random topology of the Erd\"{o}s-R\'{e}nyi model, the regular
lattice and the small-world topology among others. Such a
nomenclature is qualitative because it relies on discrete categories. By
contrast, we wish to adopt a quantitative perspective on this problem,
whereby topology is operationalized in terms of specific topological
properties such as the clustering coefficient (CC), for instance. In this perspective, 
two Erd\"{o}s-R\'{e}nyi models with identical edge
probability may display different levels
of global and local efficiencies and will therefore be considered
to have distinct topological properties. Therefore, we
distinguish between a qualitative approach based on topological
taxonony and a quantitative approach based on topological properties.
Given that generative models are latent, our quantitative
definition of topology appears better suited to the empirical study
and comparison of complex networks. 

The above definition of network topology, however, assumes that the
networks under comparison have identical numbers of vertices and
edges. When this is not the case, or when one is comparing two
populations of weighted networks, the question of whether or not these
networks have similar topological properties become arduous. Our main aim,
in this paper, is therefore to identify the situations within which
one can safely conclude that different weighted networks share the
same topological properties. In particular, we explore whether
cost-integration answers this problem. Specifically, we consider
whether cost-integration is a useful way of disentangling weighted
cost from topology. 

The paper is organized as follows. We first introduce, in section
\ref{sec:notation}, some of the
notation and basic concepts that will be used throughout the paper. In
section \ref{sec:weighted nets}, we describe the two general
families of topological measures for weighted networks, which are (i)
the weighted and (ii) cost-integrated metrics. These quantities are
first defined for a single network. The main contribution of this
paper is then reported
in section \ref{sec:comparison}, where different approaches to 
weighted network comparison are outlined, using theoretical results and simple
examples. Section \ref{sec:real data} describes an application of
these techniques to a repeated measures fMRI task investigating 
working memory. This also allows us to illustrate a Monte
Carlo (MC) sampling scheme to approximate the different measures of interest. In section
\ref{sec:discussion}, we discuss the findings of this paper in
light of the current utilization of networks in the biomedical sciences. 
Finally, we close with a set of
recommendations on how to conduct weighted network analysis in
practice and how to report the findings arising from this type of research. 
An R package entitled NetworkAnalysis
(\url{http://CRAN.R-project.org/package=NetworkAnalysis}) has been
developed that makes available the methods discussed in this paper.

\section{Network Types and Topologies}\label{sec:notation}

\subsection{Unweighted, Weighted and Fully Weighted Networks}
For clarity of exposition and consistency with the previous literature, 
we will here employ the notation used by \citet{Kolaczyk2009}. 
A comprehensive introduction to the theory of complex networks 
can be found in \citet{Newman2010}. In the following, the terms metrics and measures
will be used interchangeably to refer to a function
quantifying the topological structure of a network. Our use of the
terms metric and measure is unrelated to the mathematical definitions
of these concepts in topology and measure theory,
respectively. Similarly, we here utilize the graph-theoretical definition
of the term \ti{cost}, which is not related to its use in a probabilistic
setting. 

An unweighted undirected graph or network $G$ is formally defined as
an ordered pair $(\cV,\cE)$,
where $\cV$ is a set of vertices, points or nodes, and $\cE$
is a set of edges or connections linking pairs of nodes. Therefore
$\cE\subseteq \cV\otimes \cV$, where $\otimes$ is the Cartesian
product. The cardinality --i.e.~
the number of elements-- of $\cV$ and $\cE$ will be
referred to as $N_{V}:=|\cV|$ and $N_{E}:=|\cE|$, respectively, where
$|\cdot|$ denotes the number of elements in a set, and 
$:=$  that the left-hand-side is defined as the right-hand-side. 
Moreover, the terms network and graph will be used
interchangeably. A graph with the maximal number of edges is 
referred to as a complete or saturated graph. For a given network
$G$, we denote the corresponding saturated graph as
$G^{\op{Sat}}$. The cardinality of the edge set of $G^{\op{Sat}}$ is
denoted by $N_{I}$ to distinguish it from $N_{E}$. Here, the set
$\cI(G)$, for any graph $G$ is the set of indices of all possible
edges in $G$. That is, 
\begin{equation}
       \cI(G) := \lb (i,j) : 1 \leq i<j \leq N_{V}\rb.
       \label{eq:edge set2}           
\end{equation}
This notation for the set of indices of all possible edges in $G$ will
be useful when describing the topology of $G$ based on its shortest
paths.

Weighted undirected graphs will be denoted by the triple
$G:=(\cV,\cE,\cW)$, where $\cW(G)$ is a set of weights, whose elements are
indexed by the entries in $\cE(G)$, such that 
\begin{equation}
     w_{e_{i}} = w_{v_{j}v_{h}},
     \label{eq:weighted convention}
\end{equation}
for some edge $e_{i}:=v_{j}v_{h}$. Thus, every weighted undirected
graph will necessarily satisfy 
\begin{equation}
       N_{E} = N_{W} \leq N_{I}, 
\end{equation}
where $N_{W}:=|\cW(G)|$. The weight set populates a symmetric matrix
$\bW$, whose diagonal elements are null. Graphs that satisfy $N_{W}=N_{I}$ will be
referred to as \ti{fully weighted graphs}. 
Note that, in general, we will not draw an explicit difference between a
weighted and an unweighted network through our notation. However, which one we are referring to
should be understandable from the context. 

There are a wide range of different weighted measures of internodal
association. Our methodological development, in this paper, 
applies to any choice of association metric.
This includes correlation coefficients, partial correlations, synchronization
likelihoods and others. For simplicity, we will assume that the association weights
$w_{ij}$'s lie in the unit interval, $[0,1]$. Roughly, these
standardized weights, $w_{ij}$, can be
interpreted as the strength of the association between nodes $i$ and $j$, 
with larger values indicating a greater level of association.
Such standardization can be obtained straightforwardly, in
practice. For the case of the Pearson's correlation coefficient $r_{ij}$, for
example, the standardized weights can be defined as, 
\begin{equation}
     w_{ij} := 1-\lt(\frac{1-r_{ij}}{2}\rt).
\end{equation}
Note that the use of such a standardization of correlation
coefficients potentially leads to two major pitfalls. Firstly, 
since negative correlations are transformed into positive measures of
association, it follows that we are amalgamating different 
subsets of edges, which may play very different roles. That is, while 
subnetworks of negatively correlated vertices may reflect inhibitory
processes, subnetworks of positively correlated vertices may reflect
excitatory processes. 
Secondly, since pairs of vertices linked by
a small amount of correlation, either positive or negative, will be
transformed to take a value close to $0.5$; it follows that we may be
introducing a spurious amount of random noise in such a weighted
network analysis, as correlation coefficients close to zero are likely
to be non-significant. Our approach to weighted network analysis in
this paper, however, centres on thresholding the weighted networks of
interest and therefore does not explicitly take into account the
direction of the association.
Moreover, our focus will be on fully weighted networks, such as a standardized
correlation matrix, where all entries are greater than 0. Therefore,
in the sequel, $G$ will refer to a fully weighted graph, except when
specified otherwise. We will discuss the use and limitations of cost-integration
for non-fully weighted graphs in the discussion. 

\subsection{Classical Measures of Network Topology}
A wide range of network topological metrics have been proposed in
the literature \citep[see][for a review]{Rubinov2010}. Two types
of measures are generally of interest, which are sometimes referred to
as (i) integration metrics and (ii) specialization metrics. The
former category of topological measures quantifies a network's capacity to transfer
information globally, whereas the latter reflects a network's capacity
to transfer information locally. This distinction originated with the
work of \citet{Watts1998}, who considered the characteristic path
length (CPL), on one hand, and the clustering coefficient (CC), on the other
hand, as measures of global and local information transfer,
respectively. Although these metrics have been successfully used in a wide
range of settings, \citet{Latora2001} have introduced two analog
metrics: the global and local efficiencies, which will be more
useful in our context. These two measures
retain the interpretation of the CPL
and CC, while being applicable to a wider range of
networks. Specifically, global and local efficiency metrics can be
computed for any network, irrespective of their level of sparsity,
which is not true for CPL and CC. (That is, CPL becomes infinite when a
graph is disconnected and CC becomes undefined when a vertex has no
neighbors --that is, when a node is isolated.) Throughout this paper, and
following other authors \citep{Achard2007}, we will
therefore focus on the comparison of families of networks, whose
topologies are characterized by efficiency metrics. 

One of the remarkable aspects of global and local efficiencies is that
they can both be subsumed under the general concept of 
information transfer efficiency, which is defined for any unweighted graph
$G=(\cV,\cE)$ --connected or disconnected-- as \citep{Latora2001}, 
\begin{equation}
    E(G) := \frac{1}{N_{V}(N_{V}-1)}\sum_{i=1}^{N_{V}}\sum_{j\neq i}^{N_{V}} d^{-1}_{ij}
           = \frac{1}{N_{I}}\sum_{\cI(G)} d^{-1}_{ij},
   \label{eq:general efficiency}  
\end{equation}
where the summation over the set $\cI(G)$ is over all the pairs of
indices $(i,j)$ as in equation (\ref{eq:edge set2}), and $d_{ij}$ denotes the length of the shortest path
between vertices $i$
and $j$ in the adjacency matrix of $G$, with $d_{ij}:=\infty$ when
these two nodes are not connected.
The summation over $j\neq i$ includes all indices
between $1$ and $N_{V}$ different from $i$. The global and local
efficiencies of network $G$ are then readily derived from 
equation (\ref{eq:general efficiency}), such that
\begin{equation}
    E^{\op{Glo}}(G) := E(G), \qq\te{and}\qq 
    E^{\op{Loc}}(G) := \frac{1}{N_{V}} \sum_{i=1}^{N_{V}} E(G_{i}),
   \label{eq:efficiencies}  
\end{equation}
where $G_{i}$ is the subgraph of $G$ that includes all the
neighbors of the $i\tth$ node. That is, $\cV(G_{i}):=\lb v_{j}\in
G| v_{j}\sim v_{i}\rb$, where $v_{j}\sim v_{i}$ signifies that nodes
$i$ and $j$ are connected. By convention, we have $v_{i}\notin \cV(G_{i})$ 
\citep[see][]{Latora2001,Latora2003}. Note that both global and local
efficiencies are normalized quantities with values in the unit
interval --that is $E(G)\in[0,1]$. The global efficiency of a graph $G$
can be interpreted as the average `speed' of
information transfer between any pair of nodes in $G$, with a
high value of $E^{\op{Glo}}(G)$ indicating a high average `speed', and
therefore efficient information transfer. Similarly, the local
efficiency of a graph $G$ can be interpreted as the average global
efficiency of the $N_{V}$ subgraphs of $G$, where again a high
value for $E^{\op{Loc}}(G)$ implies efficient local information transfer,
on average.

We have used $E(G)$ to denote the efficiency metric of the unweighted graph $G$. 
This should be distinguished from the graph-theoretical concept of the
edge set, which we have denoted $\cE(G)$. Since both
quantities are functions of $G$, we have emphasized this distinction through
our notation. Note also that we will make use of the expectation
operator from probability theory, which will be denoted by
$\E[\cdot]$. For simplicity, all our development, examples and technical results
will be based on the general efficiency described in equation
(\ref{eq:general efficiency}). However, these methods could readily be extended to both
global and local efficiencies. In fact, most our discussion applies to all
topological metrics that can be computed for any level of sparsity. 
We will discuss the generalization of our results to other topological
measures in section \ref{sec:discussion}.

\subsection{Cost and Weighted Cost}
In network analysis, it is often of interest to quantify the cost or
wiring cost of an unweighted graph. In this section, we extend this concept to weighted
networks. This generalized version of cost will be termed the weighted
cost or weighted density.

The cost or density, $K:=K(G)$, of an unweighted network $G=(\cV,\cE)$ quantifies the
relative number of connections in $G$ as a proportion of the number of
edges contained in the $N_{V}$-matched saturated network
$G^{\op{Sat}}$. That is, 
\begin{equation}
     K(G):= \frac{|\cE(G)|}{|\cE(G^{\op{Sat}})|} = \frac{N_{E}}{N_{I}},
     \label{eq:cost}
\end{equation}
where $N_{I}:=N_{V}(N_{V}-1)/2$. The computation of the cost of a
network $G$ implicitly refers to the
adjacency matrix $\bA$ of that network. Hence, we can reformulate the
definition in equation (\ref{eq:cost}) by explicitly using $\bA$ as follows, 
\begin{equation}
     K(G) = \frac{\sum_{i=1}^{N_{V}}\sum_{j\neq i}^{N_{V}}
       a_{ij}}{\sum_{i=1}^{N_{V}}\sum_{j\neq i}^{N_{V}}
       a^{\op{Sat}}_{ij}} = 
       \frac{1}{N_{I}}\sum_{\cI(G)} a_{ij},
      \label{eq:cost-adjacency}
\end{equation}
where the $ a^{\op{Sat}}_{ij}$'s denote the elements of the adjacency
matrix $\bA^{\op{Sat}}$, which represents a saturated network on
$N_{V}$ nodes. 

Similarly, it will be of interest to quantify the cost of a weighted
network, which will be referred to as $K_{W}(G)$. We define it by
generalizing the relationship between an unweighted graph and its
adjacency matrix in order to apply it to weighted graphs and their association
matrices. However, to extend the concept of cost to a real-valued association
matrix, say $\bW$, we need to formalize what we mean by a \ti{saturated
weighted graph}. A natural choice is to define $\bW^{\op{Sat}}$
as a matrix of order $(N_{V}\times N_{V})$ with unit
entries. Formally, $\bW^{\op{Sat}}:=\bone_{(N_{V}\times N_{V})}$.
Using this saturated association matrix, we can now define the
cost of a weighted graph as follows,
\begin{equation}
   K_{W}(G) := \frac{\sum_{i=1}^{N_{V}}\sum_{j\neq i}^{N_{V}} w_{ij}}
                    {\sum_{i=1}^{N_{V}}\sum_{j\neq i}^{N_{V}}
                      w^{\op{Sat}}_{ij}}
            =  \frac{1}{N_{I}}\sum_{\cI(G)} w_{ij},
           \label{eq:weighted cost}
\end{equation}
where $w_{ij}^{\op{Sat}}$'s are the elements of $\bW^{\op{Sat}}$.
The non-standardized version of the cost of a weighted network 
in equation (\ref{eq:weighted cost}) was introduced by \citet{Fallani2008}.  
Thus, the weighted cost of $G=(\cV,\cE,\cW)$ is the mean
of the off-diagonal elements in $\bW$, populated by the set
$\cW$. This is reminiscent of our
starting point in equation (\ref{eq:cost}), where the same observation
can be made about unweighted networks. In the sequel, the concept of
weighted cost will be used interchangeably with the phrase
\ti{connectivity strength}. Note that depending upon which standardization one chooses, one
may obtain different types of weighted costs. In particular,
$K_{W}$ could also be standardized with respect to the number of
elements in $\cW$. This would produce a different measure. In this
paper, we will assume that the networks under consideration are fully
weighted, such that $N_{E}=N_{W}=N_{I}$, and therefore these two types
of weighted costs are equivalent.  

\section{Measures of Weighted Network Topology}\label{sec:weighted nets}
There is currently no guidance in the literature on how to quantify
the topological aspects of a weighted network. We
review here two approaches to this problem: (i) weighted, and (ii)
cost-integrated metrics of network topology. We describe and define these
two families of measures, in turn. 

\subsection{Weighted Measures}\label{sec:weighted}
A natural approach to the problem of quantifying the topology of
weighted networks is to translate unweighted measures,
such as efficiency metrics, for example, into a weighted format. 
This is a very general procedure, which has been introduced by several
authors including \citet{Latora2001} and \citet{Rubinov2010}. Weighted
versions of classical metrics commonly rely on the definition of a 
weighted shortest path. For unweighted networks, the shortest path
$d_{ij}$ between nodes $i$ and $j$ in $G=(\cV,\cE)$ is defined as the
following minimization, 
\begin{equation}
      d_{ij} := \min_{P_{kl}\in\cP_{ij}(G)} |\cE(P_{kl})|,
\end{equation}
where $\cP_{ij}(G)$ is the set of all paths between nodes $i$
and $j$ that are subgraphs of $G$. A subgraph $P_{ij}\subseteq G$ is a
path if and only if $i,j\in \cV(P_{ij})$ such that
\begin{equation}
     \cE(P_{ij}) = \lt\lb ia,ab,\ldots,yz,zj\rt\rb,
\end{equation}
where each pair of letters stands for an edge. 
One can similarly define a \ti{weighted shortest path}, $d^{W}_{ij}$, for some
weighted graph $G=(\cV,\cE,\cW)$ as follows,
\begin{equation}
    d^{W}_{ij} := \min_{P_{kl}\in\cP_{ij}(G)}\sum_{uv\in\cE(P_{kl})} f(w_{uv}), 
    \label{eq:weighted shortest path}
\end{equation}
where the weighted edge set of a path now takes the form, 
\begin{equation}
     \cW(P_{ij}) =  \lt\lb w_{ia},w_{ab},\ldots,w_{yz},w_{zi}\rt\rb,
\end{equation}
using the notational convention introduced in equation (\ref{eq:weighted convention}).
Since we have normalized the association weights, $w_{ij}$'s,
the real-valued function $f(\cdot)$ is restricted to
a map of the form $f:[0,1]\mapsto [0,1]$.  A convenient choice of $f(\cdot)$ is
the inverse function, $f(w_{ij}):=1/w_{ij}$. It now suffices to use our
chosen definition of the weighted shortest path $d^{W}_{ij}$, in order
to obtain a weighted version of the general efficiency metric 
in equation (\ref{eq:general efficiency}), which gives
\begin{equation}
     E_{W}(G) := \frac{1}{N_{V}(N_{V}-1)}\sum_{i=1}^{N_{V}}\sum_{j\neq
    i}^{N_{V}} \frac{1}{d^{W}_{ij}} = \frac{1}{N_{I}}\sum_{\cI(G)} \frac{1}{d^{W}_{ij}}.
    \label{eq:weighted definition}
\end{equation}
Note that weighted efficiency is not necessarily bounded between $0$ and
$1$. Here, we have $d^{W}_{ij}\in [\min(w_{ij}),\infty]$, regardless of
whether or not the $w_{ij}$'s have been standardized. 
The case $d_{ij}:=\infty$ may occur when there does not exist a path between $i$ and
$j$. However, since we have restricted the scope of this paper to
fully weighted matrices, where $w_{ij}\in(0,1)$ holds for every pair
of indices, it follows that $E_{W}\in\R^{+}:=(0,\infty)$. 

\subsection{Cost-integrated Measures}\label{sec:cost-integrated}
A second approach to the problem of quantifying the topology of
weighted networks proceeds by integrating the metric of interest
with respect to cost. Here, some authors have integrated over a subset
of the cost range \citep[see][for example]{Achard2007}, whereas others
have integrated over the entire cost domain \citep{He2009a}. This
second family of metrics will be referred to as cost-integrated
measures. Given a weighted graph $G=(\cV,\cE,\cW)$, the general
efficiency of equation (\ref{eq:general efficiency}) can be defined 
as follows,
\begin{equation}
     E_{K}(G) := \int\limits_{\Omega_{K}(G)}E(\gamma(G,k))p(k)dk,   
     \label{eq:cost theoretical}
\end{equation}
where cost is treated as a discrete random variable $K$, with realizations
in lower case, and $p(k)$ denotes the probability density
function of $K$. Since $K$ is discrete, it can only take a countably finite
number of values, which is the following set,
\begin{equation}
    \Omega_{K} := \lt\lb \frac{1}{N_{I}},\frac{2}{N_{I}},
           \ldots,\frac{N_{I}-2}{N_{I}},\frac{N_{I}-1}{N_{I}},1.0\rt\rb
           = :\bk,
     \label{eq:costs}
\end{equation}
where, as before, $N_{I}:=N_{V}(N_{V}-1)/2=|\Omega_{K}|$. It will be useful
to treat $\Omega_{K}$ as an ordered set, $\bk$, whose elements,
$k_{t}$'s, are arranged in increasing order and indexed by $t=1,\ldots,N_{I}$.

The function $\gamma(G,k)$ in equation (\ref{eq:cost theoretical}) is a thresholding
function, which takes a weighted undirected network and a level of wiring cost as
arguments, and returns an \ti{unweighted} network. We defer
a full discussion of $\gamma$ to Appendix A, where we describe its
definition in more detail. This function is based on the percentile
ranks of the elements of $\cW$, where tied ranks are resolved by
assigning the corresponding ordering of the elements' indices.
Since there is no prior knowledge about which
values of $K$ should be favored, we specify a uniform distribution over
$\Omega_{K}$. In equation (\ref{eq:costs}), we
have excluded the null cost for standardization purposes. Since any
edge-based topology of interest will be
zero when $K=0$, this particular value is irrelevant when comparing
different populations of networks. In example \ref{exa:toy}, we will
also see that this exclusion of the point mass at $K=0$ ensures a more satisfying
standardization of $E_{K}(G)$. 

Since $K$ is treated as a discrete random variable, we can define its
probability mass function. As no particular cost levels are favored, $K$ is given
a discrete uniform distribution, such that 
\begin{equation}
    K \sim \op{DisUnif}(\Omega_{K}),
\end{equation}
where each element of $\Omega_{K}$ has an identical probability of
occurrence, which, in our case, is equivalent to 
\begin{equation}
     p(k) = \frac{1}{|\Omega_{K}|} = \frac{1}{N_{I}}, 
\end{equation}
for every $k\in\Omega_{K}$. The theoretical integration in
equation (\ref{eq:cost theoretical})
is therefore a weighted summation over a finite set of atoms
\citep[see][]{Billingsley1995}, and may be computed as follows, 
\begin{equation}
     E_{K}(G) = \sum_{t=1}^{N_{I}}E(\gamma(G,k_{t}))p(k_{t})
              =\frac{1}{N_{I}}\sum_{t=1}^{N_{I}}E(\gamma(G,k_{t})). 
     \label{eq:cost computational}
\end{equation}
where the index $t$ runs over the elements of $\Omega_{K}$ described in
(\ref{eq:costs}). 

More generally, cost-integrated metrics can be defined with respect to
a subset of the cost regimen. This perspective on the problem of
weighted network comparison has been utilized by several authors 
\citep{Eguiluz2005,Achard2007,Supekar2009}. In our
notation, a subset of the cost levels will be indicated by an interval
of the form $[k_{-},k_{+}]\subseteq \Omega_{K}$, which refers to a
finite number of values of $K$, satisfying $k_{-}\leq k\leq
k_{+}$. Integration over that subset is then defined as
\begin{equation}
     E_{[k_{-},k_{+}]}(G) =
     \int\limits_{[k_{-},k_{+}]}E(\gamma(G,k))p(k|k_{-},k_{+})dk,
\end{equation}
where the probability mass function on $K$ is normalized with respect
to the chosen domain of integration $[k_{-},k_{+}]$, such that 
$p(k|k_{-},k_{+}) = 1/(N_{I}k_{+}-N_{I}k_{-}+1)$, for
every $k$ in that interval. The
computational formula for this generalization of equation
(\ref{eq:cost computational}) is then given by
\begin{equation}
     E_{[k_{-},k_{+}]}(G)= \frac{1}{N_{I}k_{+}-N_{I}k_{-}+1}
                  \sum_{t=N_{I}k_{-}}^{N_{I}k_{+}}E(\gamma(G,k_{t})),
\end{equation}
which follows from $N_{I}k_{l}=l$, using the definition of cost
in equation (\ref{eq:cost}). Note that the value of the conditional
probability $p(k|k_{-},k_{+})$ will be different if semi-open
intervals such as $(k_{-},k_{+}]$ are considered, instead of closed ones. This is due
to the fact that the interval of interest is over a set of discrete
values, as opposed to a subset of the real line. 
As a special case, this notation can also handle the estimation of a particular topological
metric at a single cost level, say $k_{0}$. In such cases, the interval of
interest becomes $[k_{0},k_{0}]$. Our notation makes explicit the fact
that integration over a subset of the full cost regimen, is
conditional on the choice of such a subset.

Since $K$ has been treated as a random variable and because
$E(\gamma(G,K))$ is a function of $K$, it follows that
$E(\gamma(G,K))$ is also a random variable. The integral $E_{K}(G)$ can therefore
be seen as the expectation of $E(\gamma(G,K))$ with respect to the
distribution of $K$. This probabilistic treatment of
cost-integrated metrics will be particularly helpful when
considering how to estimate these quantities, as a Monte Carlo (MC) sampling scheme
can readily be devised in order to approximate $E_{K}(G)$, when the network of
interest is too large to be computed exactly. More details
about this sampling scheme are given in Appendix A.

\section{Pros and Cons of Integrating over Cost Levels}\label{sec:comparison}
We now turn to the main question tackled in this paper: Is it useful
to integrate over the different cost levels of a particular weighted
network? In order to answer this question, we briefly consider some of
the alternatives to this approach. This consists of (i) fixing a
cutoff point, (ii) fixing
a cost regimen, (iii) integrating over all cost levels, and (iv)
directly using weighted topological metrics. Our comparison of these
four approaches is substantiated by some simple examples, synthetic data
sets, and theoretical results. For convenience, we will solely treat
the case of two weighted networks in
this section. Extensions of these ideas to the case of several populations of
networks will be discussed in section \ref{sec:discussion}. 

\subsection{Fixing a Cutoff Threshold}\label{sec:fixing cutoff}
The simplest way of comparing the topology of weighted networks
is to threshold the corresponding association matrices at a specific
value, and evaluate the resulting discrete topologies. It is instructive to
study the consequences of such a naive thresholding on two networks
with proportional association matrices, as we describe in the
following example. 
\begin{exa}\label{exa:proportional}
    Let two weighted networks $G_{1}=(\cV,\cE,\cW_{1})$ and
    $G_{2}=(\cV,\cE,\cW_{2})$, with standardized association matrices
    denoted $\bW_{1}$ and $\bW_{2}$, respectively;
    such that every $w_{ij,k}\in(0,1)$ where $k=1,2$ labels the two
    graphs under scrutiny. In addition, assume that 
    \begin{equation}
           \bW_{1}:=\alpha\bW_{2},
           \label{eq:proportionality}
    \end{equation}
    where $\alpha\in(0,1)$ is a scalar.
    That is, the association matrix of $G_{2}$ is simply proportional
    to that of $G_{1}$. Two such association matrices have been   
    discussed in the introduction and were illustrated
    in panel (a) of Figure \ref{fig:introduction}.
    Note that the relationship in equation (\ref{eq:proportionality}) implies that
    the diagonal elements of $\bW_{1}$ are not standardized to $1.0$.
    However, the topology and cost of weighted networks solely
    depend on the off-diagonal elements of such association matrices.
    Therefore, differences in the diagonal elements do not pertain
    to this discussion. Interestingly, it is easy to show that
    proportionality in association matrices implies 
    proportionality in weighted cost. Using equation (\ref{eq:weighted cost}), 
    we have
    \begin{equation}
           K_{W}(G_{2}) =
           \frac{1}{N_{I}}\sum_{\cI(G_{1})}\alpha w_{ij,1}
           =\alpha K_{W}(G_{1}),
    \end{equation}
    since $\alpha$ is applied elementwise. Therefore, $K_{W}(G_{2})>
    K_{W}(G_{1})$ as by assumption $0<\alpha<1$.  
\end{exa}

A naive approach to the problem of comparing the topologies of these two networks may
proceed by thresholding $\bW_{1}$ and $\bW_{2}$ at a particular
value, say $c\as$, as was done in the introduction. If we
compare these networks in terms of global efficiency, straightforward computation
of the two corresponding quantities shows that we necessarily have
\begin{equation}
           E(\kappa(G_{2},c\as)) \geq E(\kappa(G_{1},c\as)),
           \label{eq:efficiency inequality}
\end{equation}
for any $c\as\in[0,1]$, where $\kappa(G_{k},c\as):=\cI\lb\bW_{k}> c\as\rb$.
This follows since $G_{2}$, thresholded at $c\as$
has all the edges of $\kappa(G_{1},c\as)$, as well as
additional links owing to its weighted cost being higher. The
relationship in equation (\ref{eq:efficiency inequality}) is then deduced
from the monotonicity of the efficiency function with respect to
cost. Note that these inequalities would hold for both local and
global efficiencies, or any other topological metric, which is a monotonic
increasing function of the cost level. Therefore, example \ref{exa:proportional} has
shown that thresholding weighted graphs at a fixed cutoff point is
misleading, since graphs with higher weighted cost will tend
to be classified as having higher levels of global efficiency. This problem
can be remedied by fixing cost levels instead of cutoff points.

\subsection{Fixing a Cost Level}\label{sec:fixing cost}
A natural approach to the problem of separating cost from topology is
to choose a particular cost level. This may be a single value or a subset of the cost
regimen. Such a strategy has been adopted by several
authors \citep[see][for examples]{Eguiluz2005,Achard2007,Supekar2009}. 
One of the original justifications for conditioning over a subset of
the cost regimen was that topological metrics such as CPL or CC cannot be computed
for disconnected networks, thereby making it impossible to calculate 
these quantities for small cost levels. However, since comparable global and
local topological properties can also be measured using the efficiency metrics introduced by
\citet{Latora2001}, such problems do not arise when using these
topological metrics. We illustrate the consequences of integrating over a subset of the
range of $K$ with a real data example, where the original data has
been transformed. We have constructed a pathological case, which
shows that integrating over a subset of the cost levels can fail to
distinguish between topologically distinct weighted networks. 

\begin{exa}\label{exa:hybrid}
    We here consider a single functional connectivity matrix $\bW$,
    corresponding to the mean statistical parametric network (SPN) of a previously
    published data set, as described in section \ref{sec:real data}
    \citep{Ginestet2011a}. 
    The matrix $\bW$ was transformed in order to produce two other
    matrices with either a regular or a hybrid structure, denoted by 
    $\bW_{\op{reg}}:=F_{\op{reg}}(\bW)$ and
    $\bW_{\op{hyb}}:=F_{\op{hyb}}(\bW)$, respectively. The functions
    $F_{\op{reg}}$ and $F_{\op{hyb}}$ simply re-organize the
    position of the entries in $\bW$, as can be seen from Figure 
    \ref{fig:hybrid}. The choice of these transformations was
    constrained by the following prescriptions, 
    \begin{equation}
         \gamma\lt(G_{\op{reg}},k\pr\rt) = \gamma\lt(G_{\op{hyb}},k\pr\rt) 
         \qq\te{and}\qq
         \gamma\lt(G_{\op{reg}},k\ppr\rt) = \gamma\lt(G_{\op{hyb}},k\ppr\rt),
    \end{equation}
    for cost levels $k\pr:=0.25$ and $k\ppr:=0.75$, respectively. That is, the
    adjacency matrices corresponding to costs $k\pr$ and $k\ppr$ are
    identical for $\bW_{\op{reg}}$ and $\bW_{\op{hyb}}$. The effect of
    the functions $F_{\op{reg}}$ and $F_{\op{hyb}}$ was to create
    different layers of topological structures that vary according to
    wiring cost. The hybrid matrix was composed of alternating layers of
    random and regular topologies. Roughly, the three layers of the
    hybrid network corresponding to an hybrid association matrix
    can approximately be described as follows, 
    \begin{equation}
         \op{topology}\lt(\gamma(G_{\op{hyb}},k)\rt) = 
         \begin{cases}
              \te{Random} & \te{if }k\in[0,k\pr],\\
              \te{Regular}& \te{if }k\in(k\pr,k\ppr],\\
              \te{Random} & \te{if }k\in(k\ppr,1.0];
         \end{cases}
         \label{eq:topology regular}
    \end{equation}
    for every $k\in[0,1]$, where $k$ can only take a finite
    number of values in the unit interval. The regular
    matrix, by contrast, was built as three layers of regular
    topologies. That is,
    \begin{equation}
         \op{topology}\lt(\gamma(G_{\op{reg}},k)\rt) = \te{Regular}, 
         \label{eq:topology hybrid}         
    \end{equation}
    for every $k\in[0,1]$. The random and regular layers were 
    constructed in a standard fashion \citep[see][]{Ginestet2011a}.
    Matrices $\bW$, $\bW_{\op{reg}}$ and $\bW_{\op{hyb}}$
    corresponding to weight sets $\cW$, $\cW_{\op{reg}}$ and $\cW_{\op{hyb}}$,
    are represented in Figure \ref{fig:hybrid} with the corresponding 
    adjacency matrices resulting from different choices of cost levels.

    By construction, the weighted graphs
    $G_{\op{reg}}=(\cV,\cE,\cW_{\op{reg}})$ and $G_{\op{hyb}}=(\cV,\cE,\cW_{\op{hyb}})$
    have identical levels of general efficiency for the cost
    levels comprised in the interval $[k\pr,k\ppr]$. Therefore,
    integrating over that interval gives the same result for both graphs:
    \begin{equation}
          E_{[k\pr,k\ppr]}(G_{\op{reg}}) =
          E_{[k\pr,k\ppr]}(G_{\op{hyb}}) \doteq 0.708,
    \end{equation}
    where $\doteq$ means approximately. By contrast, the general
    efficiencies of these two networks differ substantially when 
    integrating over the full range of cost, i.e.~ $[0,1]$. This gives 
    \begin{equation}
         E_{K}(G_{\op{reg}})\doteq 0.662\qq\te{and}\qq
         E_{K}(G_{\op{hyb}})\doteq 0.679.
    \end{equation}
    This is as expected, since the hybrid network has several layers
    of random topologies, which renders it more globally efficient than
    $G_{\op{reg}}$.
\end{exa}
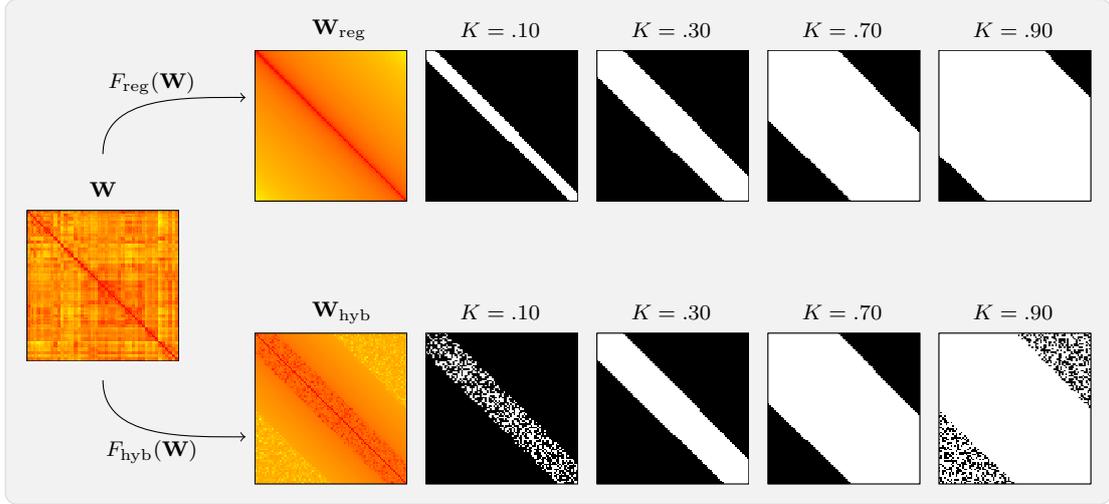
\begin{figure}[t]
  \tikzstyle{background rectangle}=[draw=gray!30,fill=gray!10,rounded corners=1ex]
  \centering
  \begin{tikzpicture}[show background rectangle,scale=.75, 
                      every text node part/.style={font=\footnotesize}]
  \draw(-1,0-.33)node[anchor=west]{\includegraphics[width=2cm]{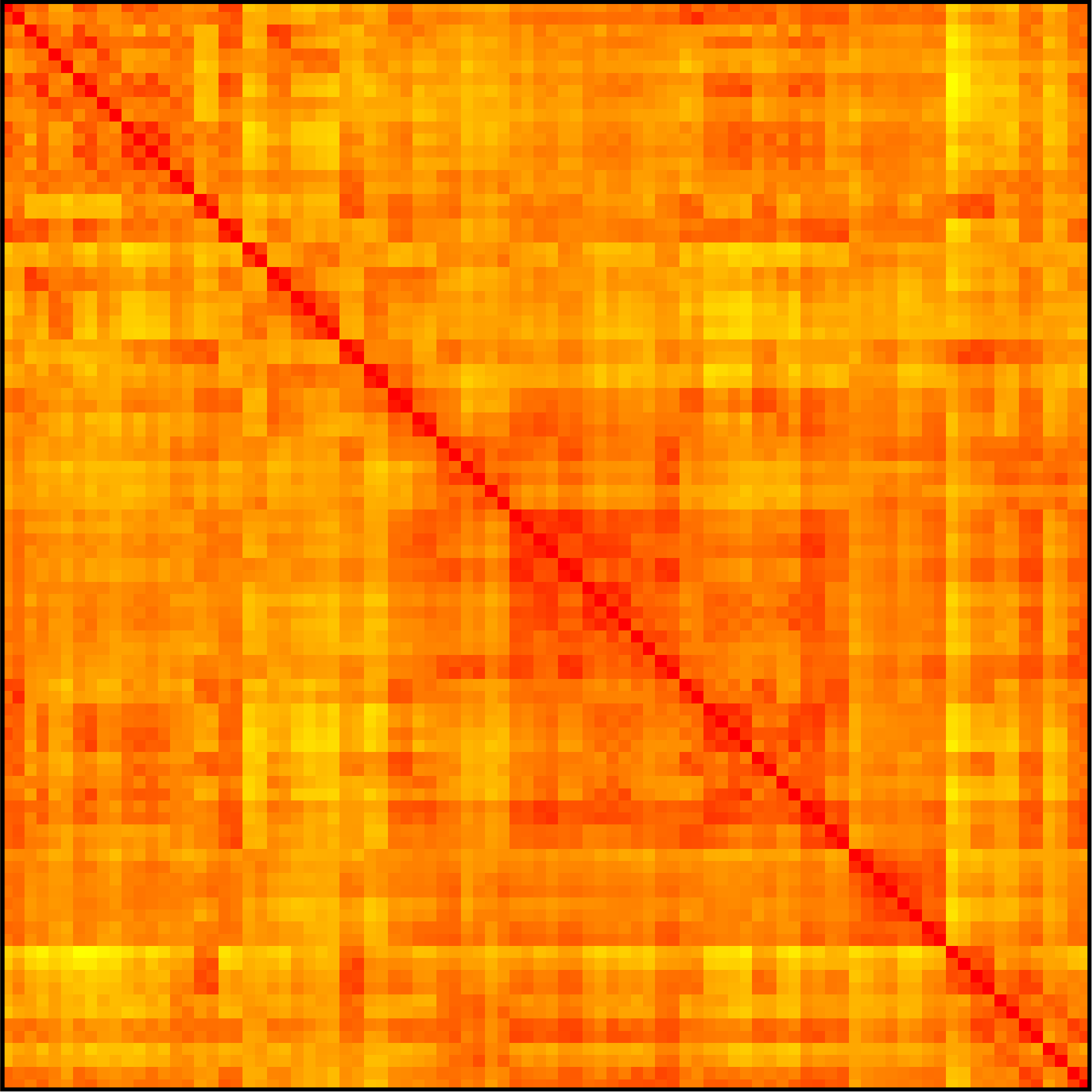}};
  \draw(0+.1,1.5+.2-.33)node[anchor=west] {$\bW$}; 
  \draw[->](0.5,2)to[out=90,in=180]node[above]{$F_{\op{reg}}(\bW)$}(3,3);
  \draw[->](0.5,-2)to[out=-90,in=180]node[below]{$F_{\op{hyb}}(\bW)$}(3,-3);

  \draw(3,+2.5)node[anchor=west]{\includegraphics[width=2cm]{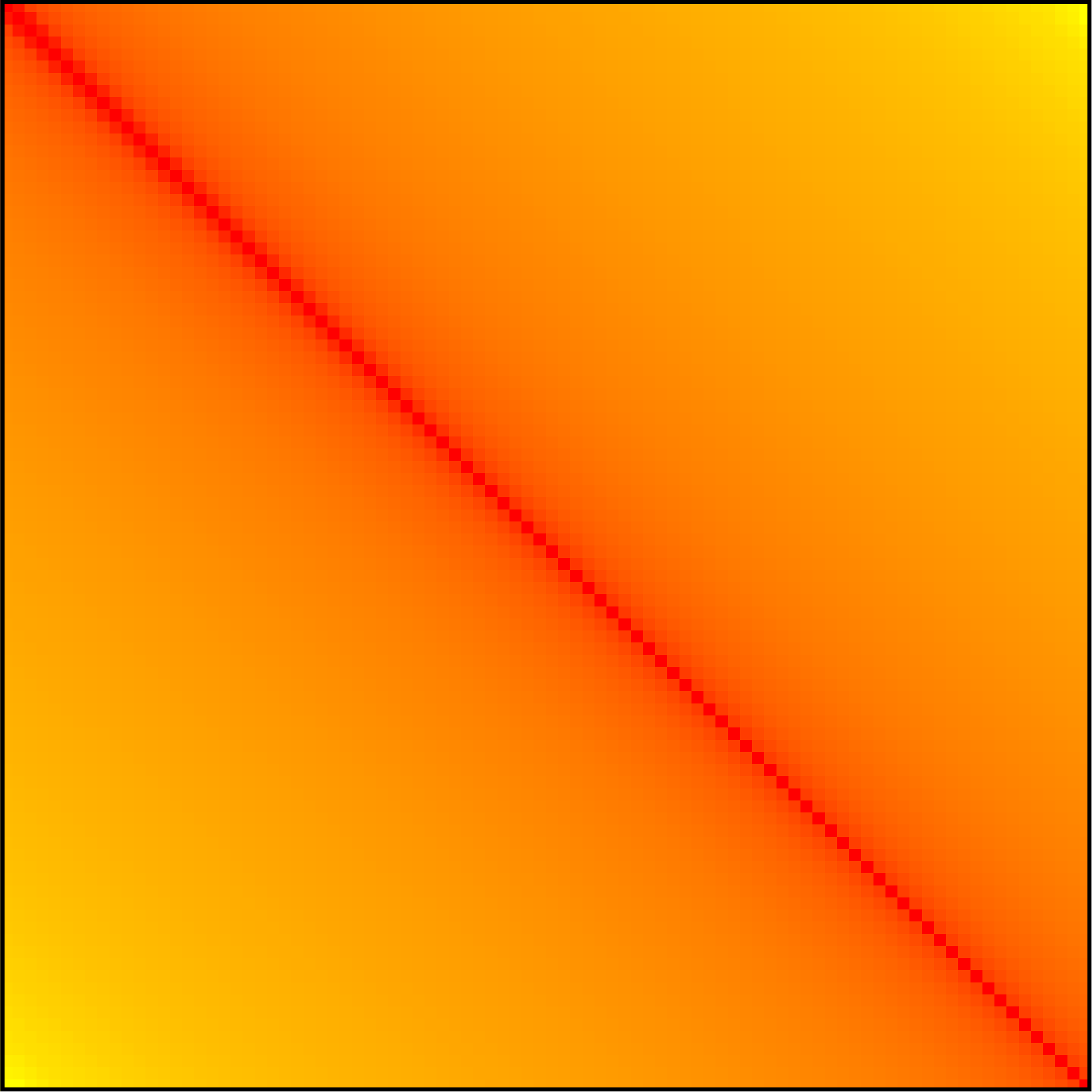}}; 
  \draw(6,+2.5)node[anchor=west]{\includegraphics[width=2cm]{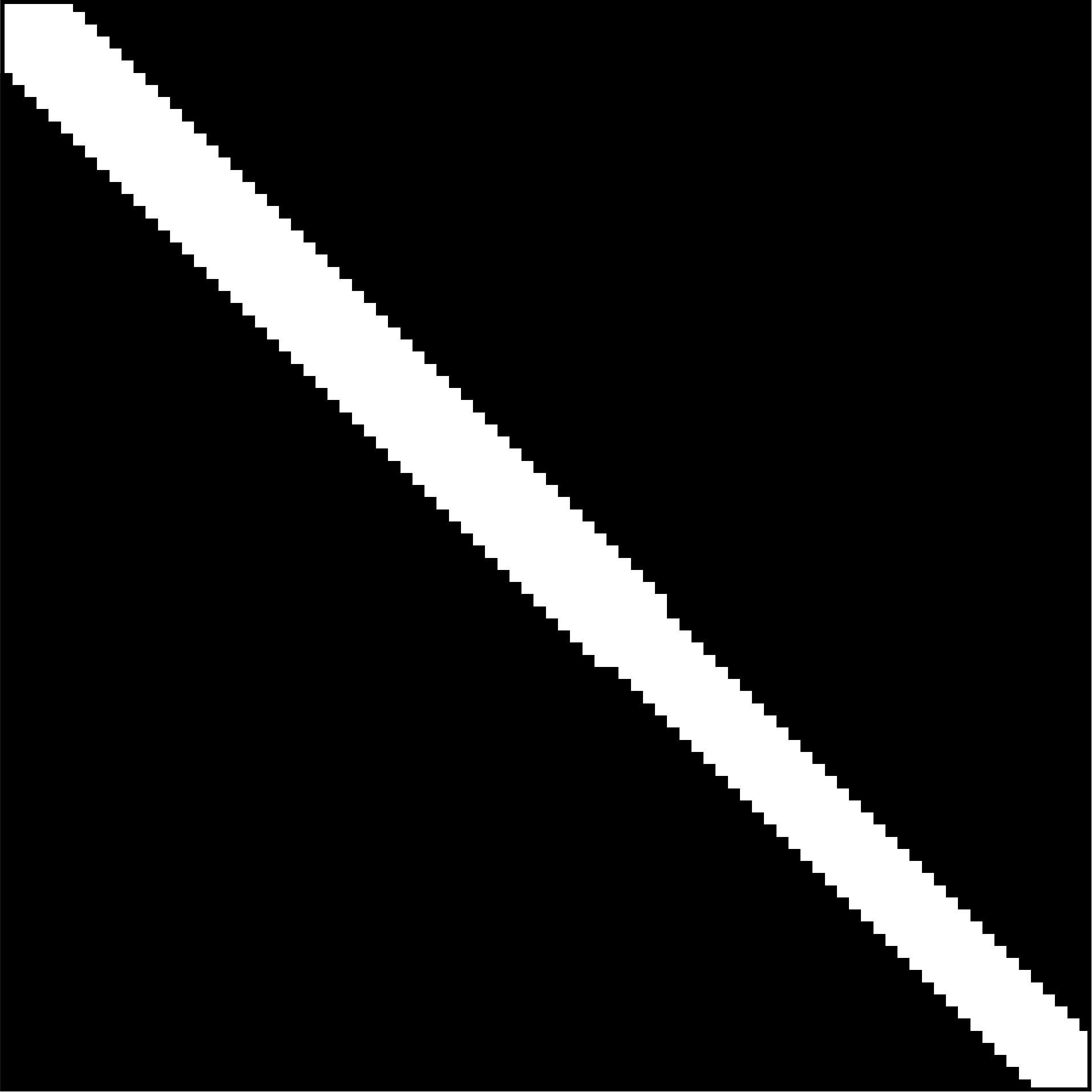}}; 
  \draw(9,+2.5)node[anchor=west]{\includegraphics[width=2cm]{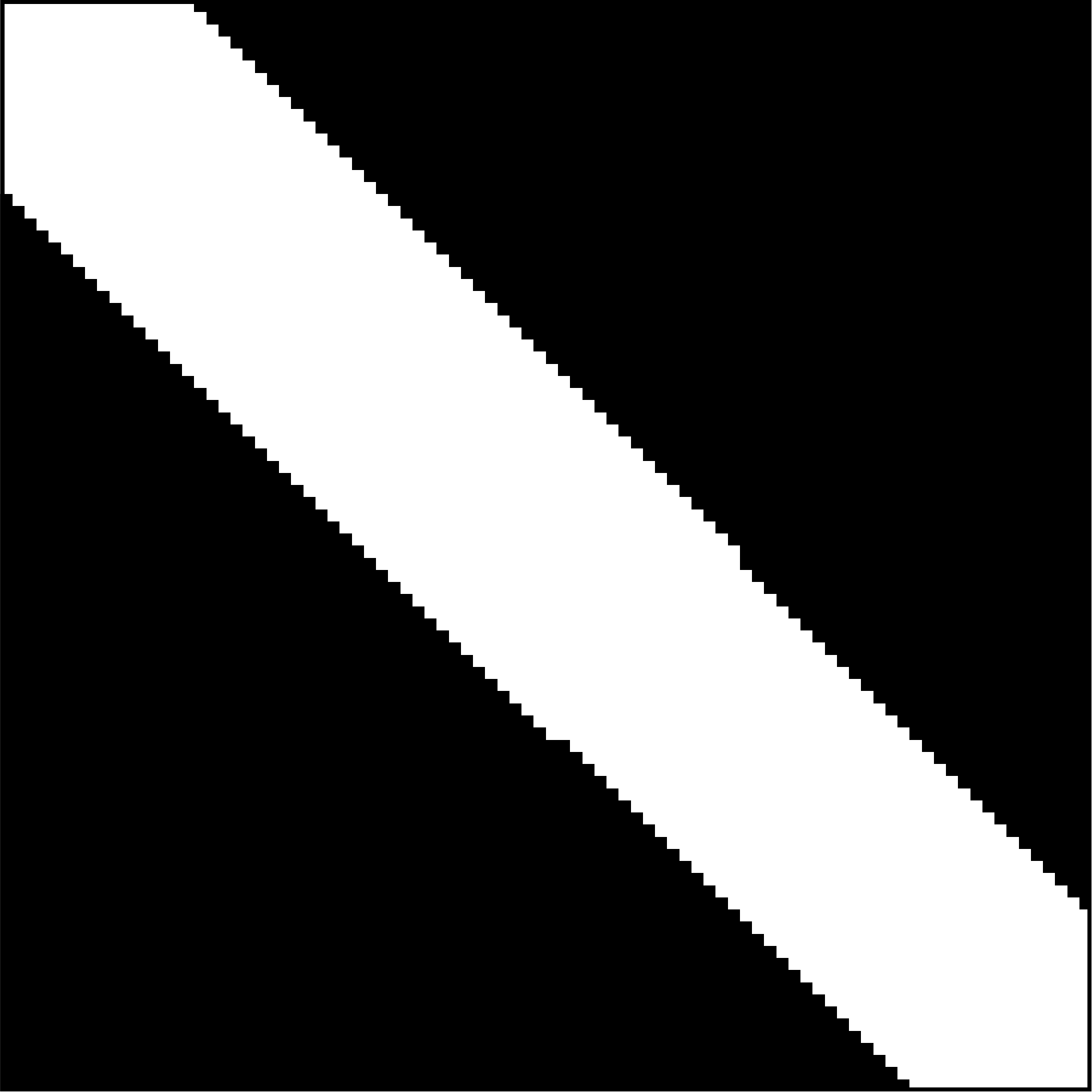}}; 
  \draw(12,+2.5)node[anchor=west]{\includegraphics[width=2cm]{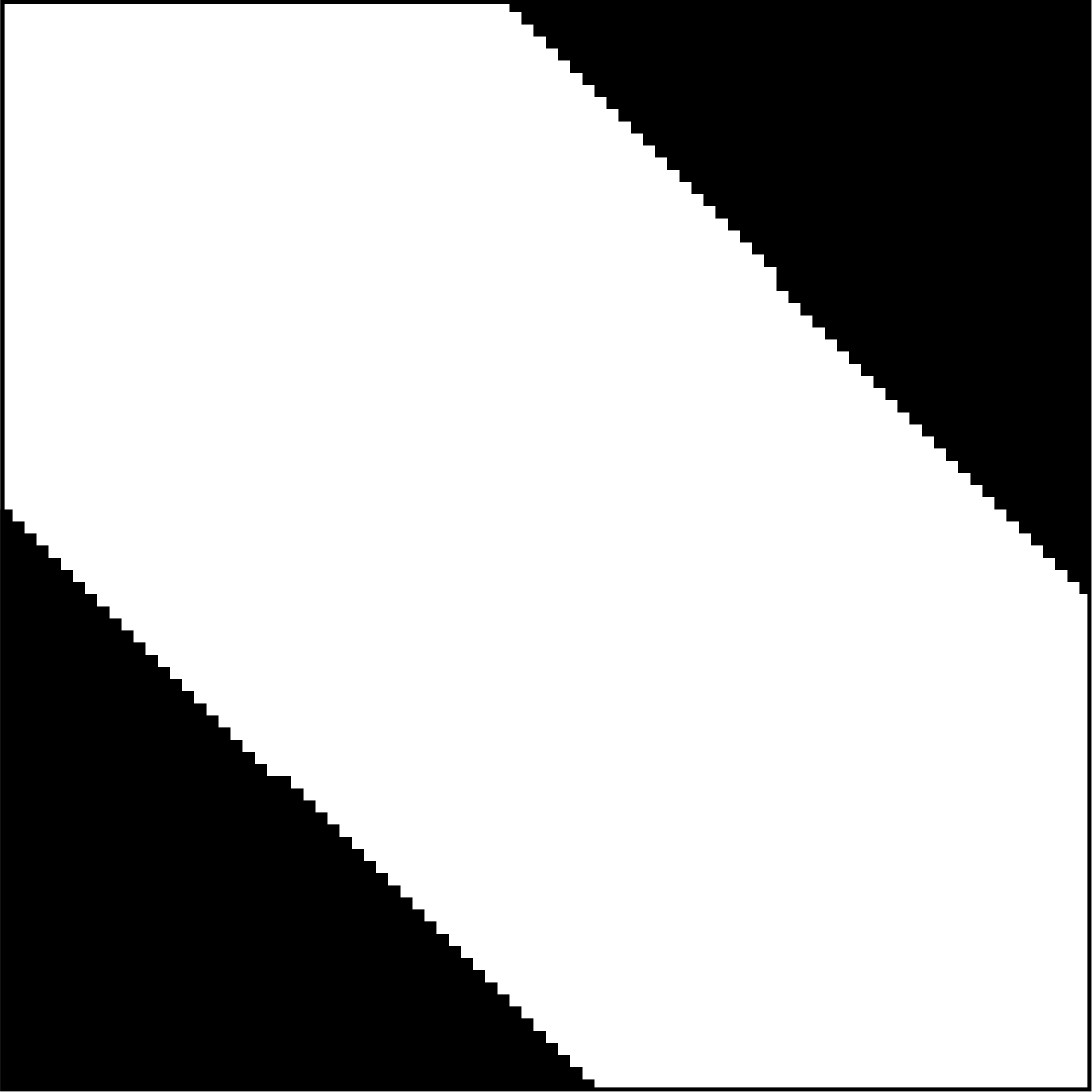}}; 
  \draw(15,+2.5)node[anchor=west]{\includegraphics[width=2cm]{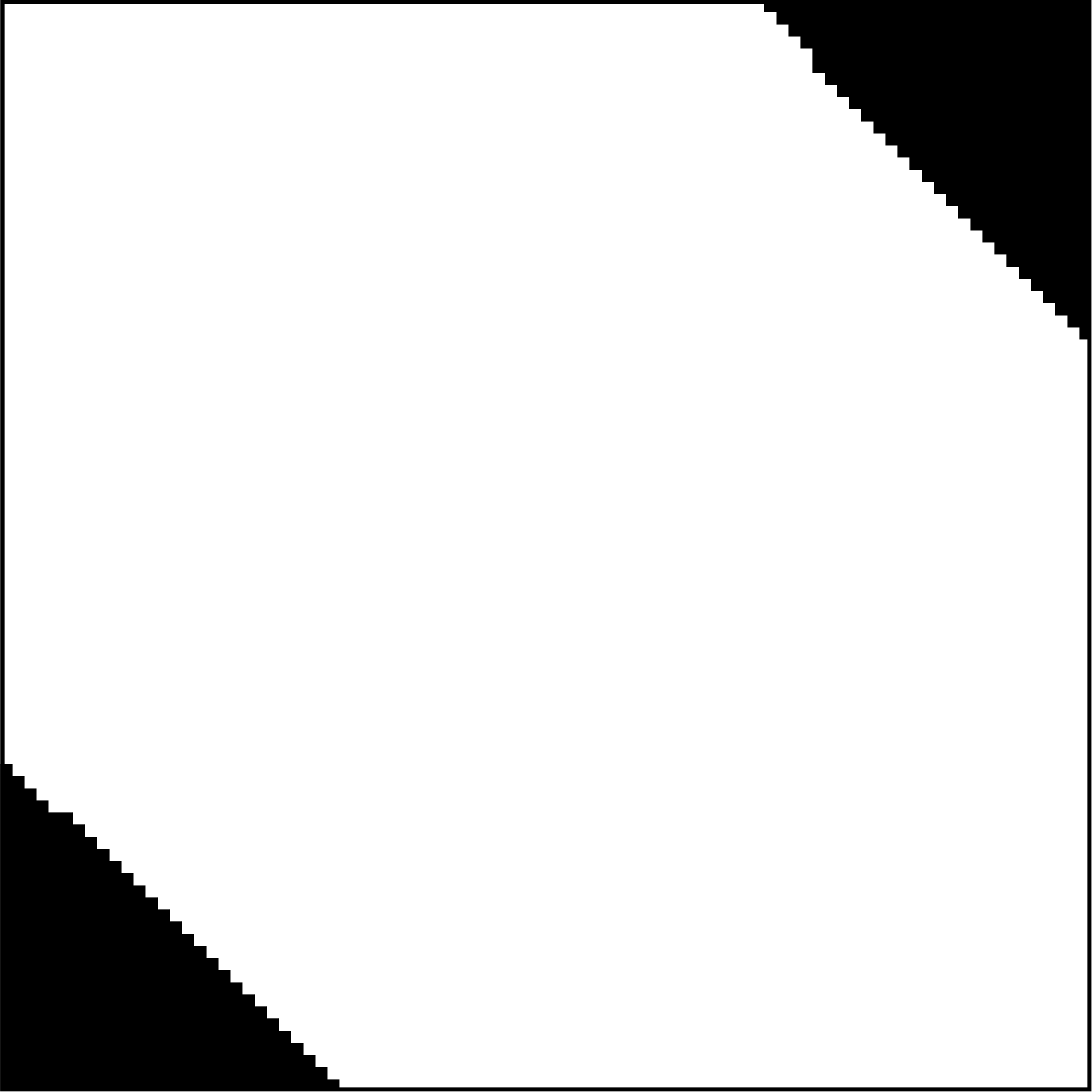}}; 
  \draw(4,4+.2)node[anchor=west]{$\bW_{\op{reg}}$}; 
  \draw(7-.4,4+.2)node[anchor=west]{$K=.10$}; 
  \draw(10-.4,4+.2)node[anchor=west]{$K=.30$}; 
  \draw(13-.4,4+.2)node[anchor=west]{$K=.70$}; 
  \draw(16-.4,4+.2)node[anchor=west]{$K=.90$}; 

  \draw(3,-2.5)node[anchor=west]{\includegraphics[width=2cm]{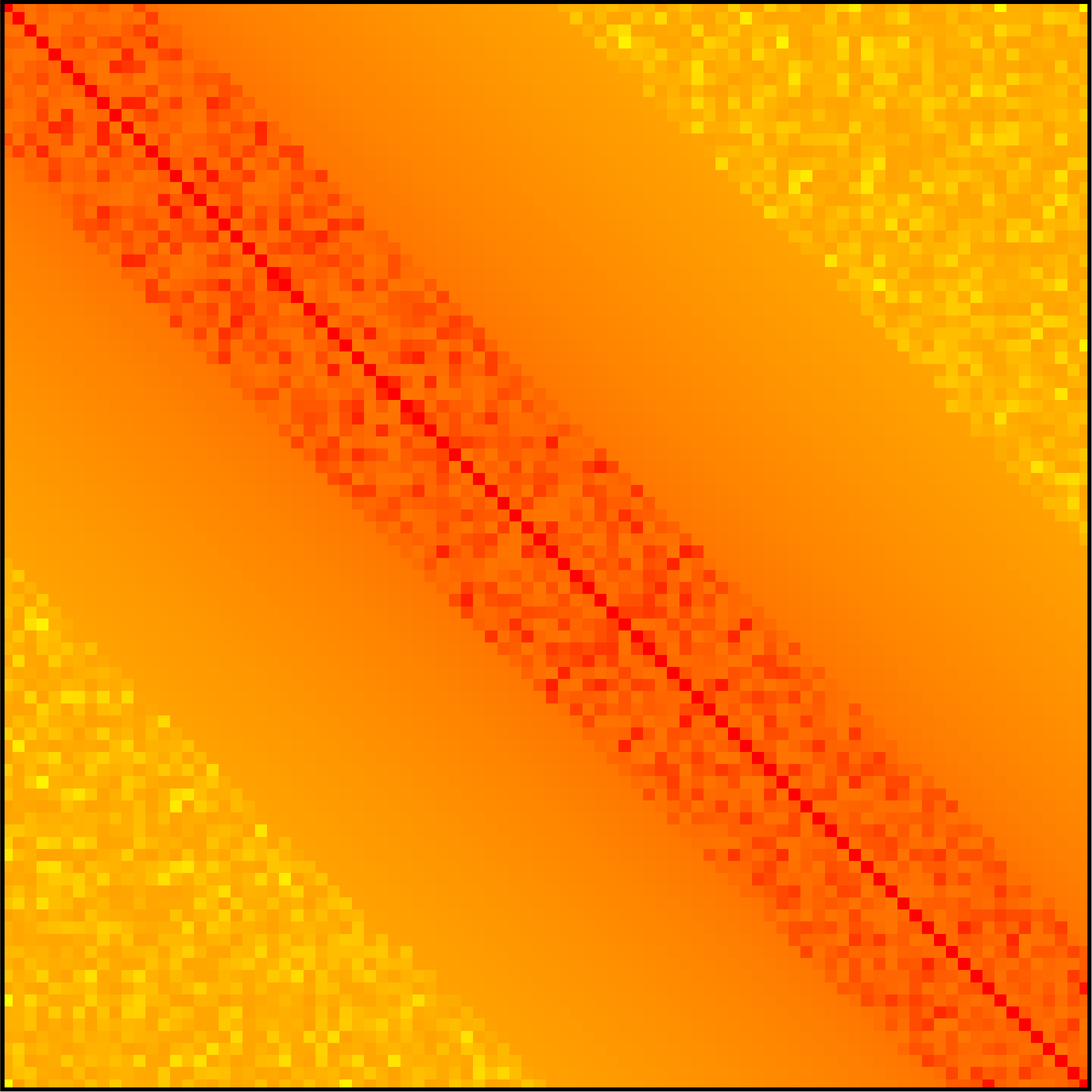}}; 
  \draw(6,-2.5)node[anchor=west]{\includegraphics[width=2cm]{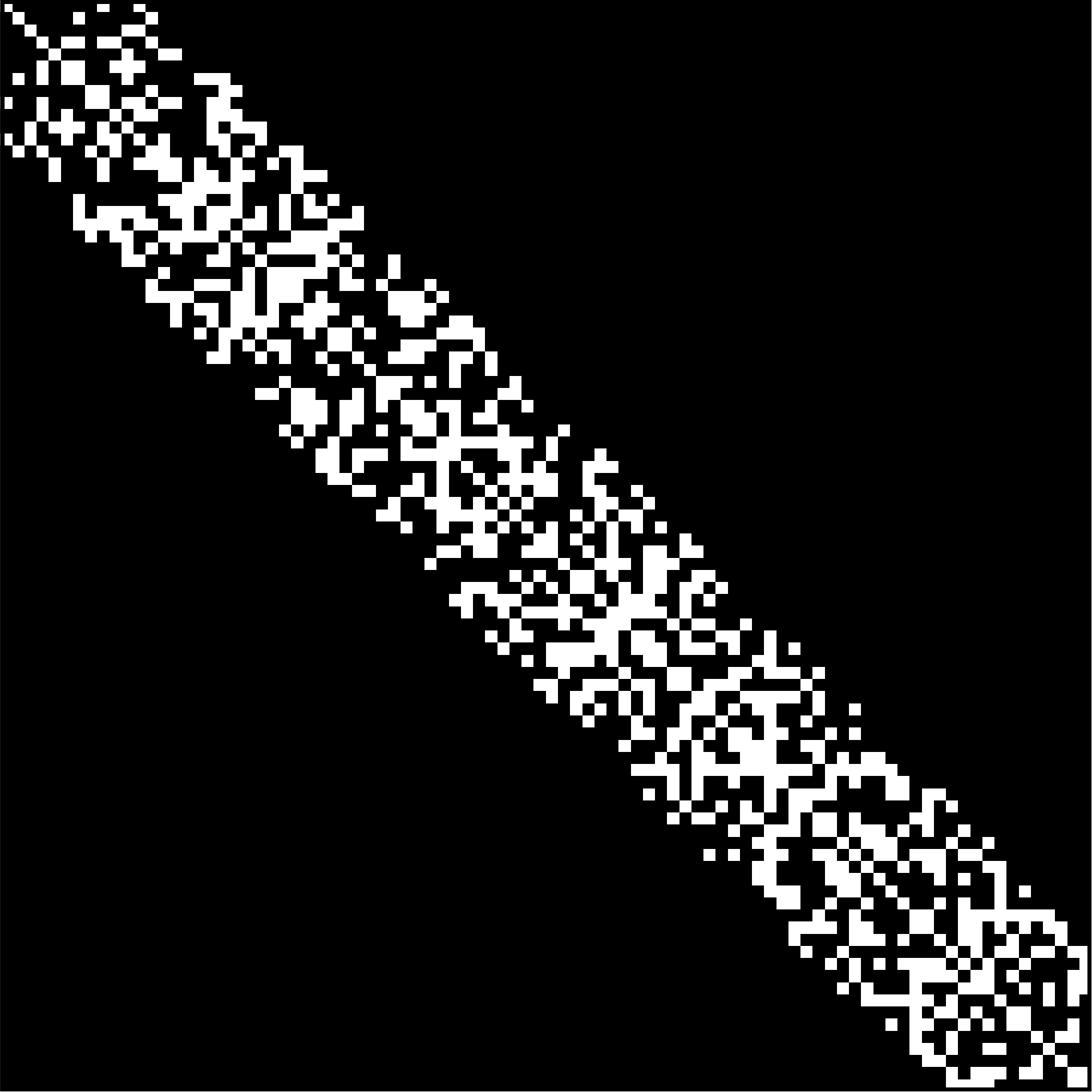}}; 
  \draw(9,-2.5)node[anchor=west]{\includegraphics[width=2cm]{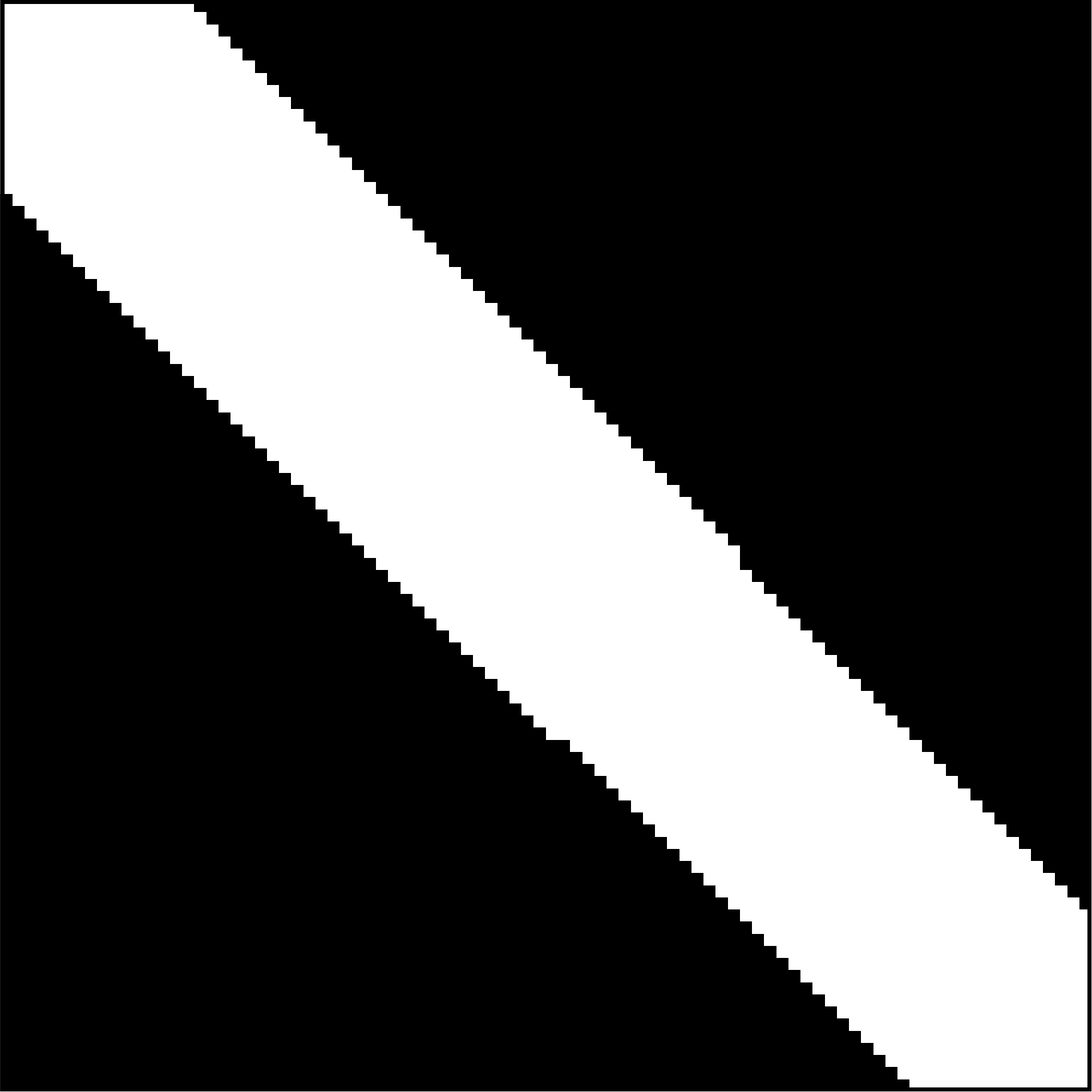}}; 
  \draw(12,-2.5)node[anchor=west]{\includegraphics[width=2cm]{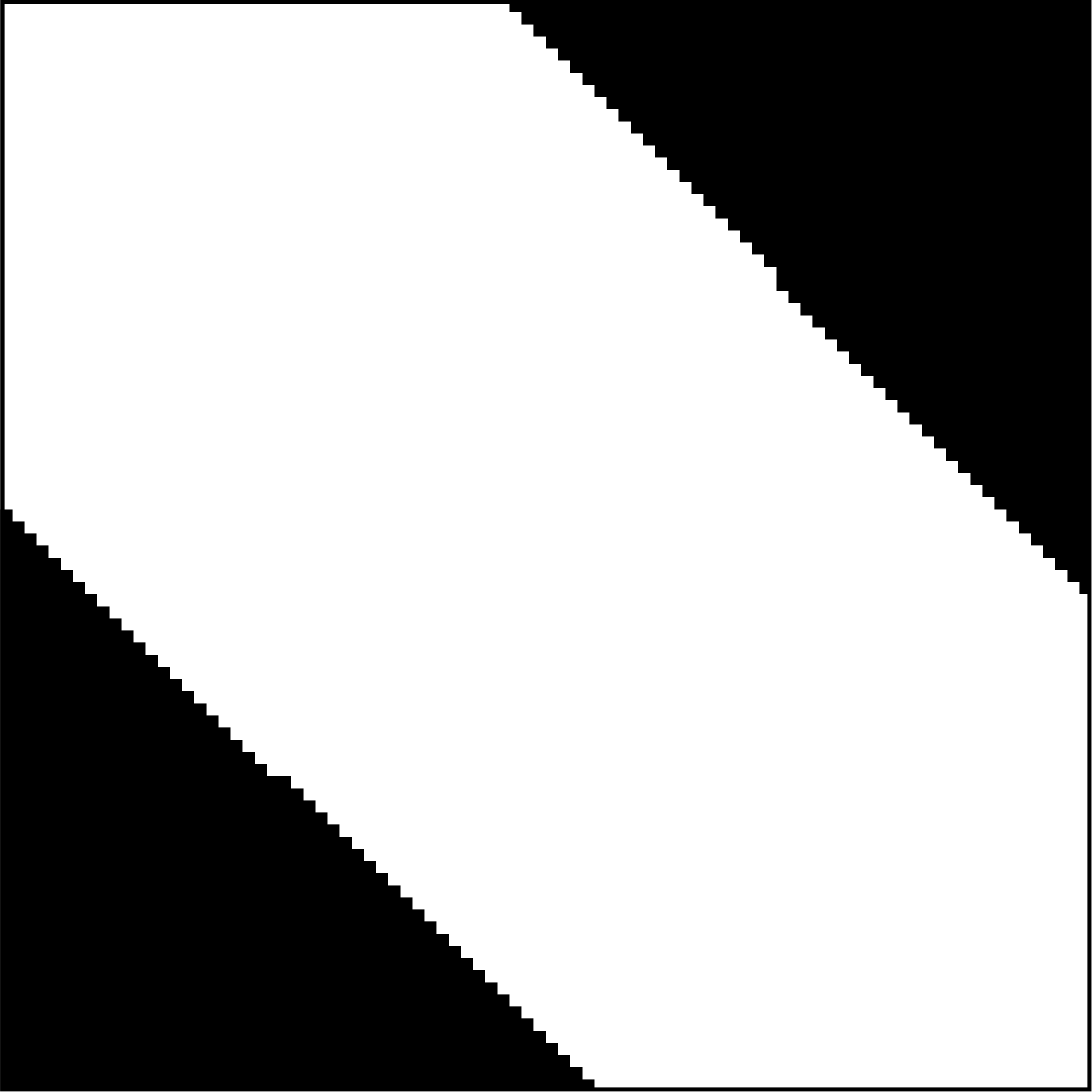}}; 
  \draw(15,-2.5)node[anchor=west]{\includegraphics[width=2cm]{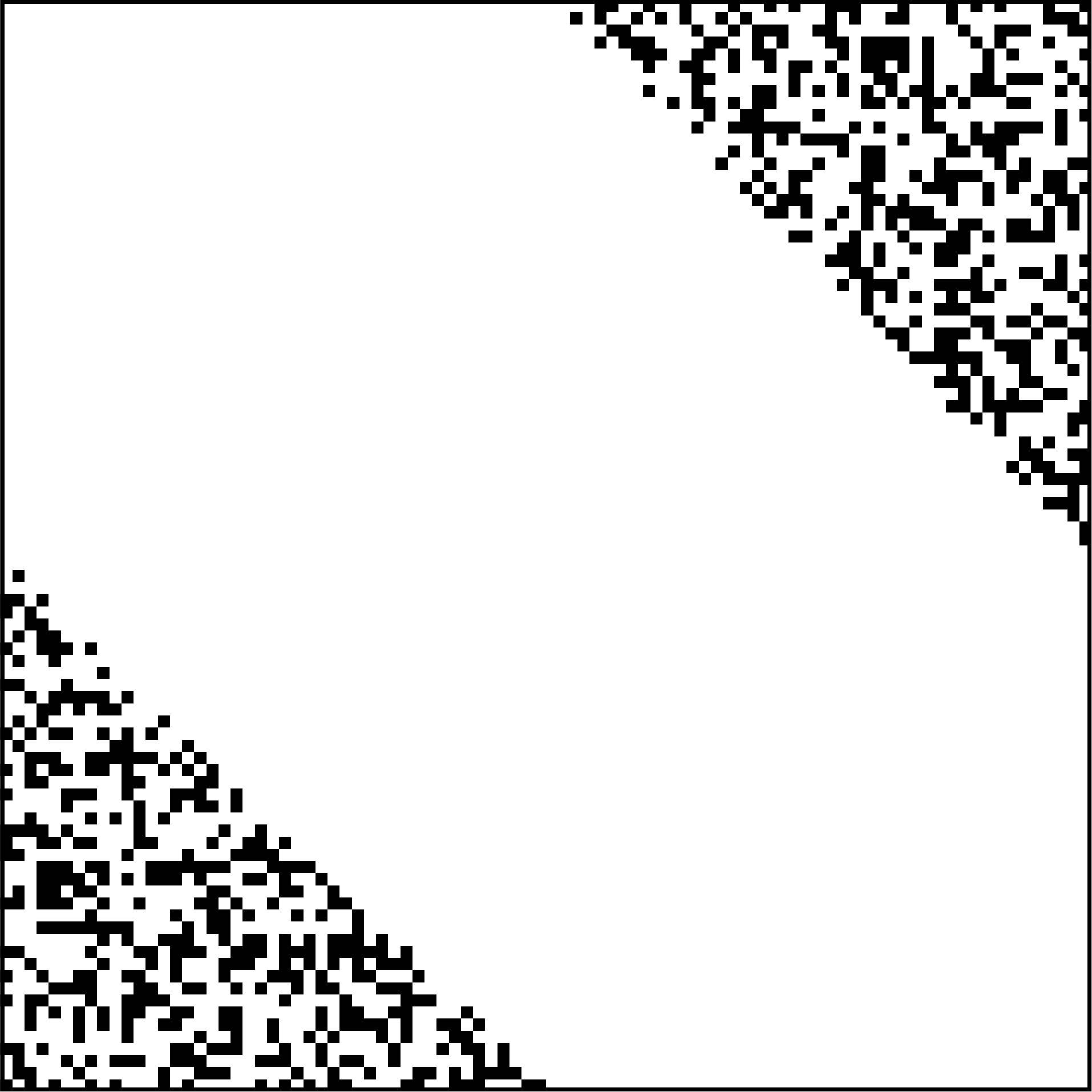}}; 
  \draw(4,-1+.2)node[anchor=west]{$\bW_{\op{hyb}}$}; 
  \draw(7-.4,-1+.2)node[anchor=west]{$K=.10$}; 
  \draw(10-.4,-1+.2)node[anchor=west]{$K=.30$}; 
  \draw(13-.4,-1+.2)node[anchor=west]{$K=.70$}; 
  \draw(16-.4,-1+.2)node[anchor=west]{$K=.90$}; 
  \end{tikzpicture}
  \caption{Simulation framework for the counterexample in section
    \ref{sec:fixing cost}. The small-world correlation matrix $\bW$ is
    transformed into a regular and a hybrid matrix, denoted
    $\bW_{\op{reg}}$ and $\bW_{\op{hyb}}$. The regular matrix exhibits 
    a lattice-like topology throughout its cost range, whereas
    $\bW_{\op{hyb}}$ consists of alternating topological layers of
    random and regular structures. The entries in both matrices have
    been arranged in decreasing order from the diagonal, to facilitate
    visualization.
    \label{fig:hybrid}}  
\end{figure}
Example \ref{exa:hybrid} illustrates the problems associated with integrating over a
subset of the cost regimen. By doing so, we are potentially omitting
substantial topological differences between the networks of interest
at other cost levels. The difference in $E_{K}$ between $G_{\op{reg}}$
and $G_{\op{hyb}}$ reported in that counterexample may
not appear very large. However, these two
networks could have represented the mean networks of two populations of
interest. Providing that the pool of subjects is sufficiently large,
such topological differences could be found to be statistically
significant. By contrast, comparison
of these two networks on the basis of the full cost regimen
yielded answers, which were exactly identical, thus nullifying any
statistical test of group differences. Naturally, this example could
have been constructed in the opposite direction in order to show that
networks that seem to differ topologically for some cost subsets are,
in fact, identical when integrating over the full cost regimen. 

Fixing a cost level or a subset of the cost
regimen therefore suffers from two main problems. Firstly, the arbitrariness of
the choice of a specific cost subset will generally be difficult to justify from
either a theoretical or a practical perspective. Secondly, as we have
illustrated with example \ref{exa:hybrid}, considering only a subset
of the cost potentially omits topological differences, which are
solely visible at other cost levels. Thus, any network analysis using this strategy
can only draw conclusions that are \ti{conditional} on the choice of
cost subset, and this dependence should be made explicit when
reporting the results of such analyses. 
Nonetheless, fixing a particular cost subset 
successfully satisfies one of our desiderata, which was to 
disentangle differences in cost from differences in topology. That is,
weighted networks' topologies can be compared irrespective of cost
differences, by conditioning on some subset of the cost levels.
This invariance property will be made mathematically more precise in
the next section. 

\subsection{Integrating over Cost levels}\label{sec:integrating cost}
From a statistical perspective, the problem of isolating topology from
connectivity strength may be reformulated as evaluating topological
differences while `controlling' for cost, where these two quantities are treated
as random variables. A natural starting point is to consider
weighted networks whose association matrices are proportional to each
other, as in the ensuing example.

\begin{exa}\label{exa:toy}
    As a simple example, consider the following problem. Let two weighted
    networks $G_{1}=(\cV,\cE,\cW_{1})$ and $G_{2}=(\cV,\cE,\cW_{2})$, be characterized by the following
    standardized association matrices: 
    \begin{equation}
        \bW_{1}:= \begin{bmatrix} 0.0 & w_{12,1} \\ & 0.0\end{bmatrix},\qq\te{and}\qq
        \bW_{2}:= \begin{bmatrix} 0.0 & w_{12,2} \\ & 0.0\end{bmatrix},
        \label{eq:toy}
    \end{equation}
    where we assume that $w_{12,1}$ and $w_{12,2}$ are comprised in
    the open interval $(0,1)$. Here, there are only two levels of
    cost, $K\in\lb 0,1\rb$. Trivially, $G_{1}$
    and $G_{2}$ can therefore be shown to exhibit identical general efficiency for these
    two cost levels. Since our proposed formula for cost-integrated
    topological measures in equation (\ref{eq:cost computational})
    does not include the null cost, we simply have $\Omega_{K}=\lb 1\rb$,
    which implies that both graphs attain the maximal level of
    cost-integrated efficiency. That is, 
    \begin{equation}
         E_{K}(G_{1}) = E_{K}(G_{2}) = 1.0.
         \label{eq:toy equality}
    \end{equation}
    This simple example serves as a justification for our exclusion of the null
    cost from the set $\Omega_{K}$ in equation
    (\ref{eq:costs}). Including the null cost would result in 
    $E_{K}=0.5$ for these two basic networks, which does not appear
    satisfying. Crucially, the equality in (\ref{eq:toy equality}) does not
    depend on the relationship between $w_{12,1}$ and
    $w_{12,2}$. That is, differences in weighted cost 
    have no impact on cost-integrated topology. 
    We now return to the case studied in example \ref{exa:proportional} 
    in order to elucidate the exact effect of cost-integration.
\end{exa}
\setcounter{exa}{0}  
\begin{exa}[Continued]\label{exa:proportional2}
    In this example, we considered two networks with proportional
    association matrices, satisfying $\bW_{1}:=\alpha\bW_{2}$. 
    An application of the cost-integrated metrics described in equation 
    (\ref{eq:cost computational}) to the networks of this example
    gives the following equalities, 
    \begin{equation}
           E_{K}(\bW_{1}) = E_{K}(\alpha\bW_{2}) = E_{K}(\bW_{2}). 
    \end{equation}
    That is, when integrating with respect to the cost levels, we are evaluating the
    efficiencies of $G_{1}$ and $G_{2}$ at $N_{I}$ discrete
    points. At each of these points, the efficiency of the two networks will
    be identical, because $\bW_{1}$ is proportional to $\bW_{2}$ and
    therefore the same sets of edges will be selected. 
    Thus, $G_{1}$ and $G_{2}$ have identical cost-integrated efficiencies. 
\end{exa}

The equalities derived in these two examples can be shown 
to hold in a more general sense. The invariance of 
cost-integrated efficiency turns out to be true for any monotonic
(increasing or decreasing) transformation of the association matrix
and applies to any topological metric that takes an
unweighted graph as an argument, as formally stated in the following
result. 
\begin{pro}\label{pro:monotonic}
   Let a weighted undirected graph $G=(\cV,\cE,\cW)$. 
   For any monotonic function $h(\cdot)$ acting elementwise on a
   real-valued matrix $\bW$, corresponding to the weight set $\cW$,
   and any topological metric $E$,
   the cost-integrated version of that metric, denoted $E_{K}$, satisfies
   \begin{equation}
          E_{K}(\bW) = E_{K}(h(\bW)),
   \end{equation}
   where we have used the association matrix, $\bW$, as a proxy
   notation for graph $G$.
\end{pro}
A proof of this result is provided in Appendix B. It relies on the
idea that the evaluation of a weighted network solely depends on the
ranking of the off-diagonal elements of $\bW$ (i.e. the ranking of
the elements in $\cW$), and that the ranks of a set
of values are independent of a monotonic transformation of these
values. Note that the arguments used in Appendix B do not rely 
on the definition of $E$. Therefore, proposition
\ref{pro:monotonic} is true for any cost-integrated
topological metrics --i.e.~ a metric originally defined in a discrete
setting for an unweighted graph, and integrated with respect to cost,
when applied to a weighted network. Note also that proposition
\ref{pro:monotonic} only holds for any level of sparsity in $G$ if the
thresholding function $\gamma(G,k)$ used in the computation of a
cost-integrated metric preserves the original ordering of elements in $\cW$ with tied
ranks, using their indices. In general, however, sparse networks may better be dealt with,
in this context, by adjusting the size of the integration domain. 

Proposition \ref{pro:monotonic} encapsulates both the advantages and
limitations of cost-integrated topological metrics. Two weighted networks,
whose topologies are roughly identical at every cost level will be
given identical scores under this family of metrics, irrespective of
cost differences. Cost-integrated
metrics are therefore successful at winnowing topology from
connectivity strength. 
Another singular advantage of this approach is that we obtain a
measure, which is invariant under 
any normalization or standardization of the original data. That is,
any functions that simply rescale or shift the association weights, in
order to ensure that they are comprised in the unit interval, for
instance, will have no effect on the value of the cost-integrated
topological measures. 

However, proposition \ref{pro:monotonic} also demonstrates
the limitation of such an approach. One can easily see that such
cost-integration will potentially mask some cost-specific topological
differences, as illustrated in example \ref{exa:hybrid}. In addition,
when cost-integrated topological metrics are used for network
comparison, this requires that the sizes of the weight sets of different networks
are identical. Similarly, the presence of multiplicities in the ranks
of the weights may also cause problems, as this would artificially induce a random
topological structure, since weights with equal ranks would be randomly allocated to
different cost levels. We will further discuss these limitations in
the conclusion of this paper. 

\subsection{Using a Weighted Metric}\label{sec:weighted metric}
A seemingly natural way of amalgamating connectivity strength and topological
characteristics is by directly considering weighted topological
metrics, such as the weighted global efficiency, $E_{W}$, introduced in equation
(\ref{eq:weighted definition}). Unfortunately, we here prove 
that such an approach suffers from a serious limitation, which could
potentially dissuade researchers from using this particular type of metrics.
With the next proposition, we show that in a wide range of
settings, the weighted efficiency is
simply equivalent to the weighted cost of the graph of interest. 
\begin{pro}\label{pro:weighted-cost}
   For any weighted graph $G=(\cV,\cE,\cW)$, whose weight set is
   denoted by $\cW(G)$, if we have
   \begin{equation}
        \min_{w_{ij}\in\cW(G)} w_{ij} \geq \frac{1}{2}\max_{w_{ij}\in\cW(G)}w_{ij},
   \end{equation}
   then 
   \begin{equation}
          E_{W}(G) = K_{W}(G).
   \end{equation}
\end{pro}
This result can be proved by contradiction, as demonstrated 
in Appendix E. The hypothesis in proposition
\ref{pro:weighted-cost} may at first appear relatively
stringent. However, it will encompass a wide range of experimental
situations. For the real data set described in example
\ref{exa:hybrid}, the difference between $\max w_{ij}$ and $2\min w_{ij}$
is close to, but not exactly zero. However, we nonetheless have
$E_{W}=K_{W}$, for that example. Thus, the added
value of using the weighted efficiency will, in general, be
questionable since there exists a strong relationship between this
topological measure and a simple average of the edge
weights. 

These theoretical results and associated counterexamples have
therefore highlighted the limitations of various approaches to the
problem of disentangling differences in cost from differences in
topology. As a result, when comparing several populations of
networks, we recommend the reporting of differences in weighted
costs and differences in cost-integrated topological
measures. We illustrate this approach with a re-analysis of a
previously published fMRI data set. 
\begin{figure}[t]
  \footnotesize
    \centering
    \tb{Sagittal $\overline{\op{SPN}}_{j}$}\\
    \vspace{.1cm}
    \ti{Sup.}\hspace{2.5cm}
    \ti{Sup.}\hspace{2.5cm}
    \ti{Sup.}\hspace{2.5cm}
    \ti{Sup.}\\
    \includegraphics[width=2.7cm]{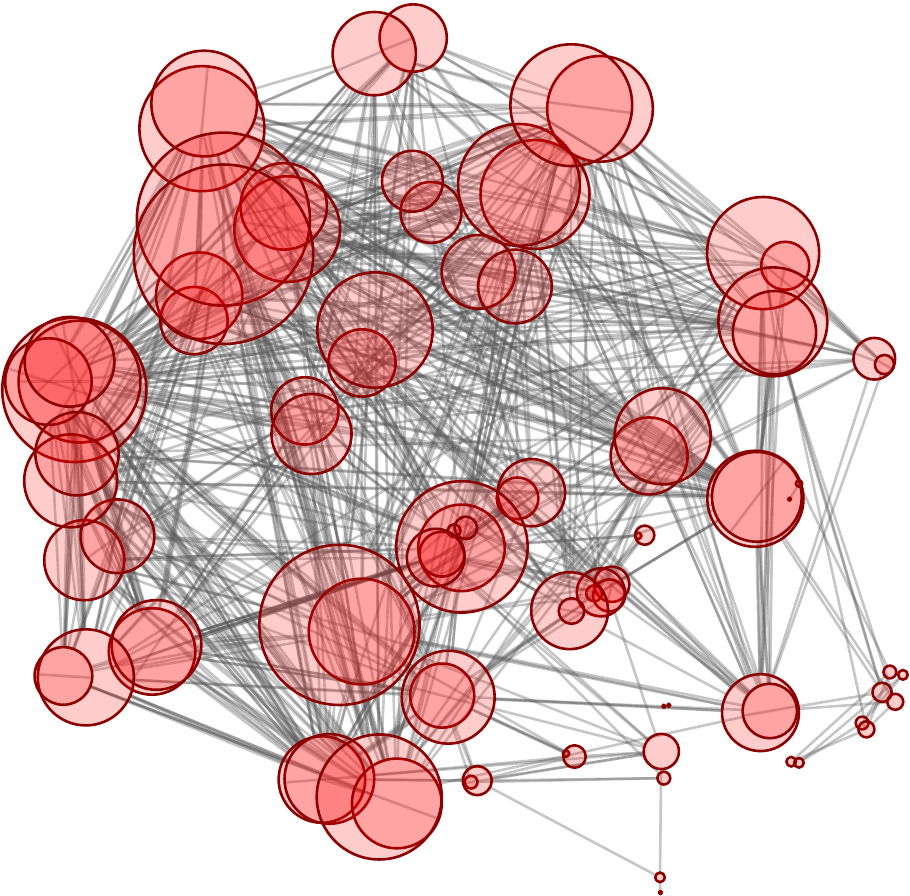}\hspace{.3cm}
    \includegraphics[width=2.7cm]{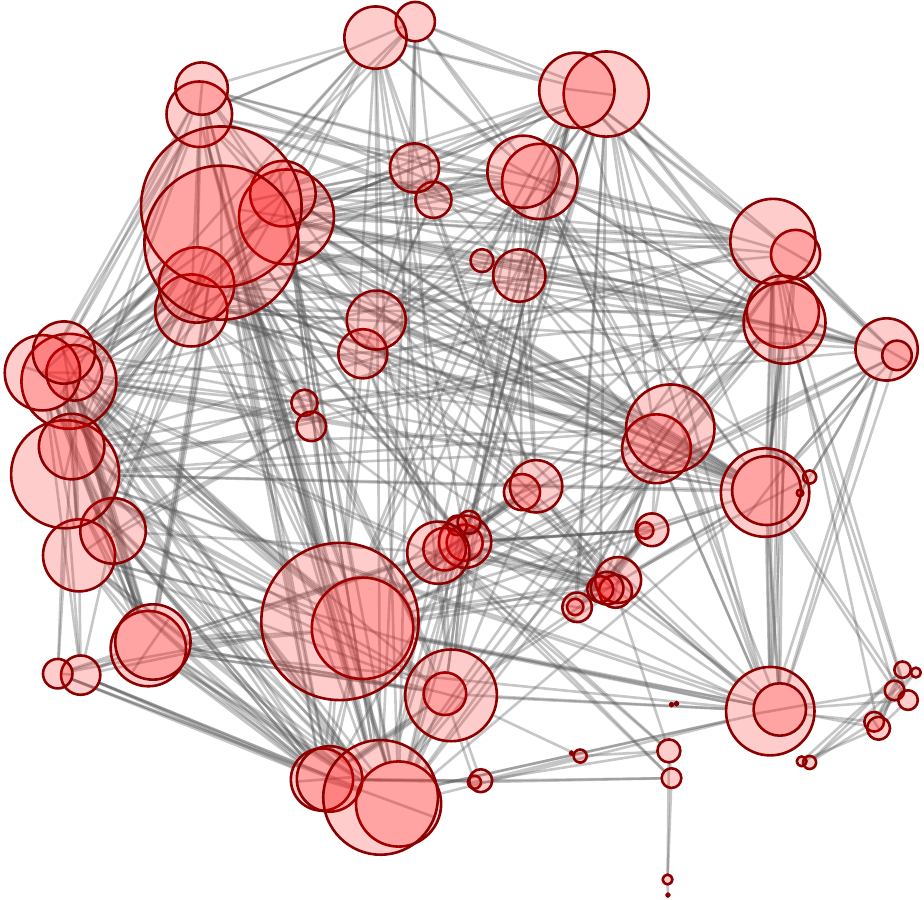}\hspace{.3cm}
    \includegraphics[width=2.7cm]{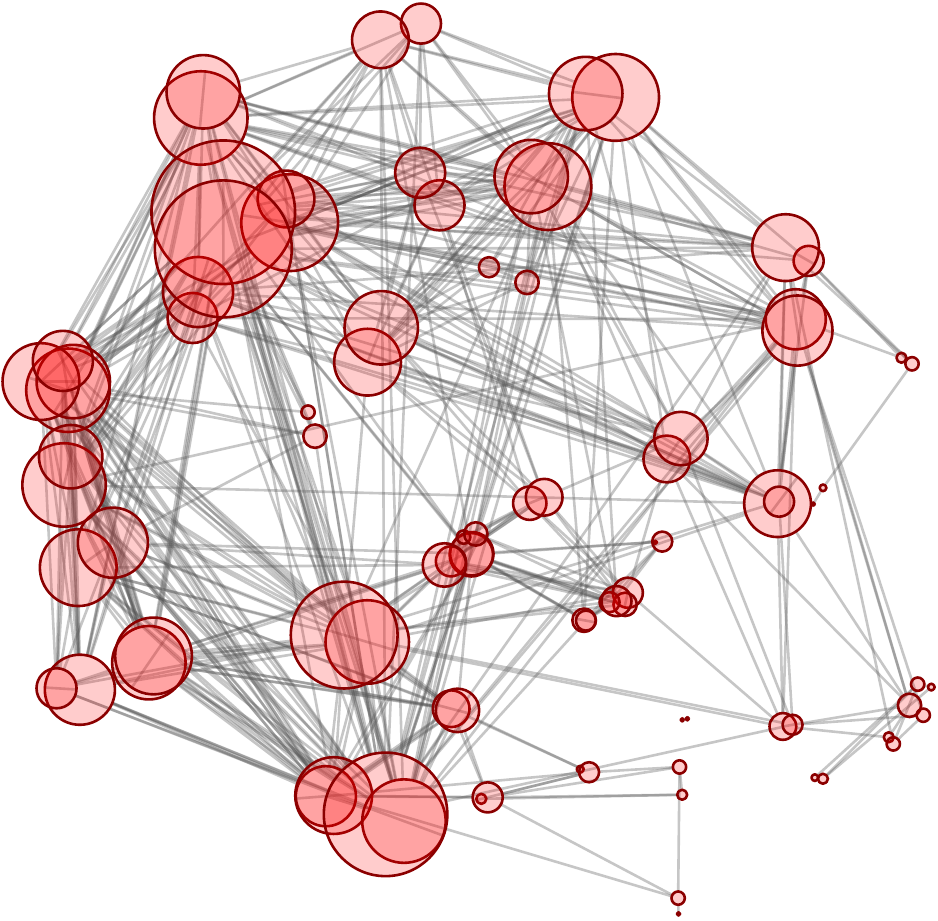}\hspace{.3cm}
    \includegraphics[width=2.7cm]{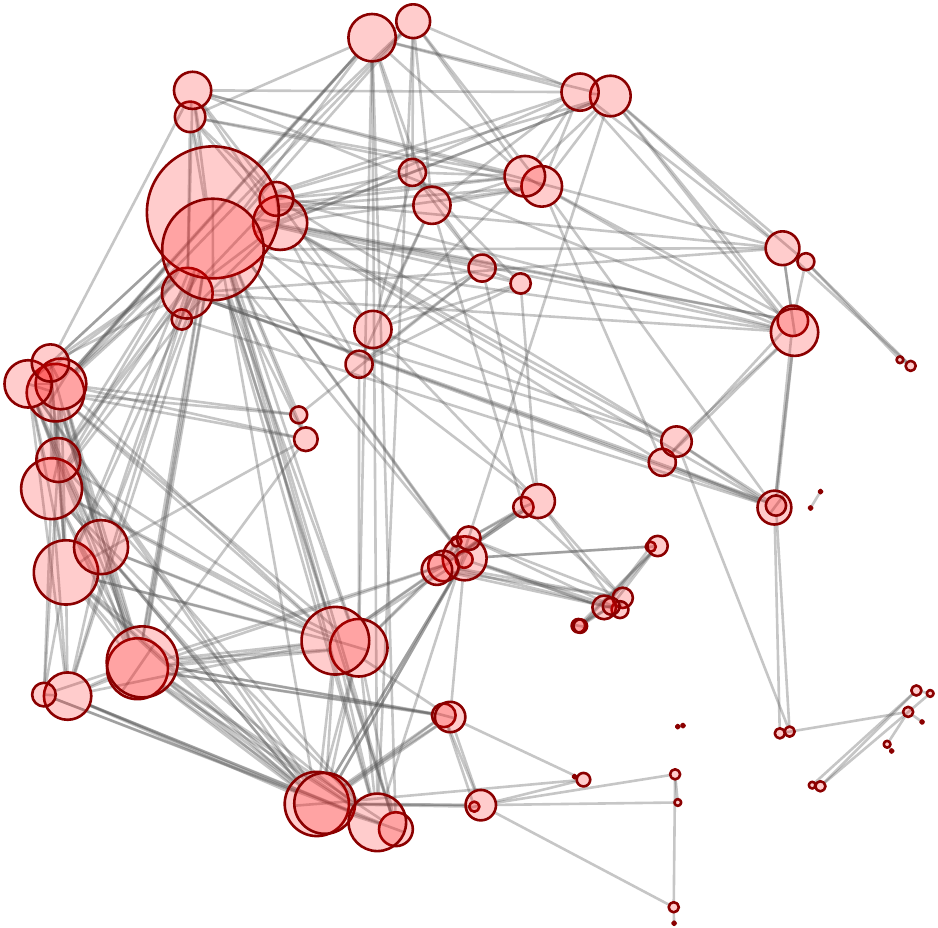}\\
    \ti{Inf.}\hspace{2.5cm}
    \ti{Inf.}\hspace{2.5cm}
    \ti{Inf.}\hspace{2.5cm}
    \ti{Inf.}\\
    \vspace{.5cm}    
    \tb{$0$-back}\hspace{2.3cm}
    \tb{$1$-back}\hspace{2.3cm}
    \tb{$2$-back}\hspace{2.3cm}
    \tb{$3$-back}
    \caption{Mean Statistical Parametric Networks
      ($\overline{\op{SPN}}_{j}$) over the 4 levels of the $N$-back
      task, in the sagittal plane, based on wavelet coefficients in the 0.01--0.03Hz frequency band, with
      FDR correction (base rate $\alpha_{0}=.05$). Locations of the nodes correspond to
      the stereotaxic centroids of the corresponding cortical regions. The
      inferior--superior orientation axis is indicated in italics.
      The size of each node is proportional to its degree.
      \label{fig:net_array}}
\end{figure}

\section{$N$-back Working Memory Data Set}\label{sec:real data}
In this section, we illustrate our theoretical results 
with a previously analyzed data set of a working
memory task based on functional Magnetic Resonance Imaging (fMRI)
data \citep{Ginestet2011a}. In particular, we use this data set
for testing our proposed MC sampling procedure and for comparing
a graph's weighted cost with its cost-integrated and
weighted global efficiencies. 

\subsection{Description}
\citet{Ginestet2011a} considered topological changes in functional
brain networks under different levels of cognitive load. 
Here, we solely give a cursory description of
the experimental procedure used in this study and refer the reader to
the original paper for the full technical details. 
\citet{Ginestet2011a} constructed networks on the basis of fMRI data gathered from 43 healthy
adults undergoing a working memory task known as the $N$-back
paradigm. In this experiment, subjects were shown one letter every two
seconds, and were asked to monitor the stimuli, in order to indicate
by the push of a button whether the current letter was identical to the
one presented $N$ trials previously, where $N=\lb 1,2,3 \rb$. A control or
null condition was also included, the $0$-back task, which consisted of
simply indicating whether the current letter was an X.
In this experiment, the subject-specific fMRI images were
parcellated into 90 regions of interest using the Anatomical
Automatic Labelling (AAL) template \citep{Tzourio-Mazoyer2002}.
The BOLD time series were averaged for each AAL region. These
regional mean time series were then wavelet decomposed. Wavelet
coefficients in the low frequency range (0.01-0.03Hz) were selected
for the main network analysis \citep[see also][for a similar
analysis of fMRI data]{Achard2006}. Since the $N$-back paradigm
contains four experimental levels, we decomposed these time series
into blocks corresponding to each $N$-back condition. As each
condition was repeated more than once, these blocks were then
concatenated. Note that this sequence of processing steps involving
wavelet decomposition immediately followed by block concatenation was
studied by \citet{Ginestet2011a} using simulated data, and was 
not found to bias the results of the final network analysis. 

Vertices in these subject-specific functional networks were
chosen to be the 90 AAL regions, and the edges were constructed by
computing pairwise correlations between each condition-specific time
series of wavelet coefficients. The results of this construction can
be summarized using Statistical Parametric Networks (SPNs), as
illustrated in Figure \ref{fig:net_array} \citep[see][for
details]{Ginestet2011a}. SPNs are estimated using 
a mass-univariate approach, where the edges in a population of
subject-specific networks are tested for significance using a
mixed-effects model, and then thresholded using the
false discovery rate \citep{Benjamini1995,Nichols2003}.
SPNs can be constructed using functions made freely available through
the R package NetworkAnalysis 
(\url{http://CRAN.R-project.org/package=NetworkAnalysis}).
From Figure \ref{fig:net_array}, one can observe that the connectivity strength 
(i.e.~ weighted cost or averaged correlation coefficient) of
the functional networks in each condition tend to diminish as
subjects experience greater cognitive load. 

\subsection{Monte Carlo (MC) Estimation}
\begin{figure}[t]
  \includegraphics[width=15cm]{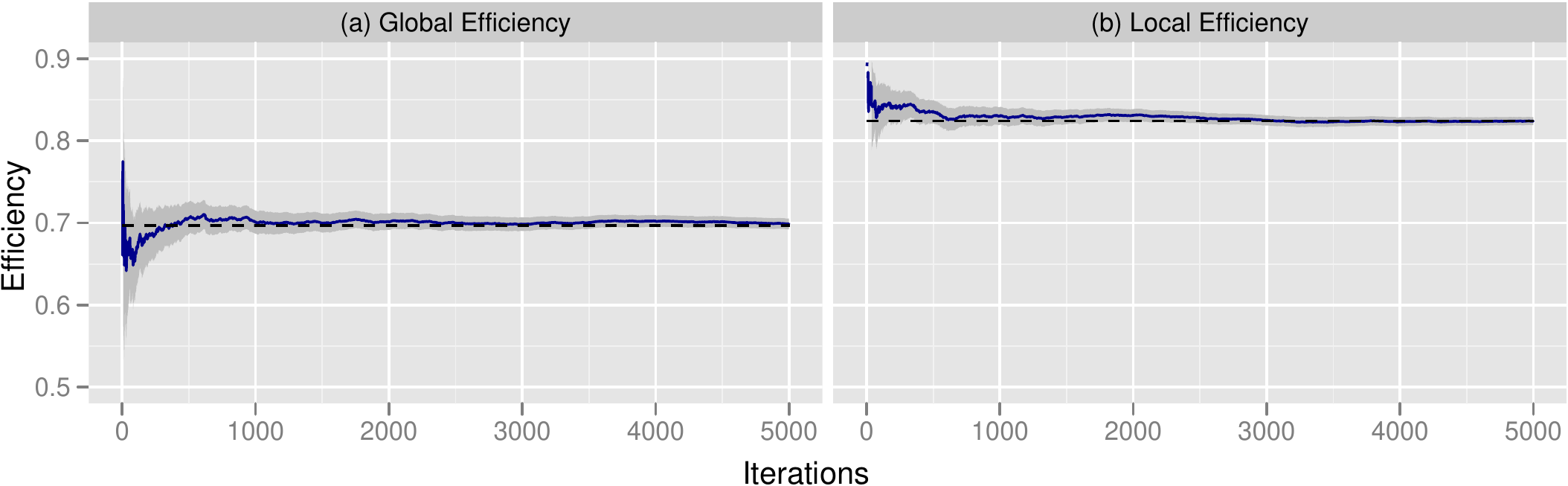} 
  \caption{Running means of Monte Carlo (MC) estimates for 
    cost-integrated global and local efficiencies in panel (a) and
    (b), respectively, for the $3$-back network described in
    example \ref{exa:hybrid}. The grey ribbon represents the
    variability of these estimators at each $m=1,\ldots,5000$, using
    twice the MC standard error. That is, $E^{(m)}_{K}\pm
    2\sig^{(m)}_{K}$ for both global and local efficiencies. The dashed lines
    indicate the exact value of $E_{K}$.
    See appendix A for details. \label{fig:mc R}}
\end{figure}

A full description of the theory supporting MC estimation in this
context is provided in Appendix A. MC techniques are here used to
speed up the computation required when estimating our proposed
cost-integrated measures. 
Figure \ref{fig:mc R} shows the convergence of
$\overline{E}^{(m)}_{K}$ to $E_{K}$, for a medium-sized weighted network derived
from fMRI data on the working memory task described in example
\ref{exa:hybrid}. The results are provided for
both global and local efficiencies. Each plot in Figure \ref{fig:mc R}
shows the running mean plus or minus twice the running MC standard error, which
are defined for the cost-integrated efficiencies, as 
$\overline{E}^{(m)}_{K}$ and $(v^{(m)}_{K})^{1/2}$, respectively, 
where $m=1,\ldots,5000$. (See Appendix A for details.)
In Figure \ref{fig:mc R}, we also report the exact values of $E_{K}$
using formula (\ref{eq:cost computational}) by dashed lines. 

In all the cases studied, the MC
estimates compared favorably with the exact integrals after
approximately a quarter of the number of computations required for the exact
calculations. That is, the exact derivation of $E_{K}$ 
necessitates $N_{I}=4005$ evaluations of the global or local
efficiency. By contrast, MC estimates based on about
$1000$ samples appear to provide reasonably good approximations of these
quantities, as indicated by the small MC standard error. This
constitutes a non-negligible computational gain. The
MC standard error, which is derived as a by-product of these
computations could then be used as an indicator of the uncertainty
associated with these estimates in a Bayesian hierarchical model,
where uncertainty is propagated from the data to the population's
parameters of interest. 

A simple alternative to MC averaging, in our context, would be 
to construct a mesh of the unit interval and to 
approximate the desired integral by a weighted sum of the values of the
topological metric of interest at the midpoints of that mesh. The
latter method is generally referred to as the Gauss-Kronrod quadrature
formula \citep[see chapter 2 of][for a review of integration
methods]{Minka2001}. While this method is very efficient for simple
functions, it becomes rapidly unwieldy
for complex ones, as it requires an increasingly refined mesh to
ensure good interpolation. Moreover, since the Gauss-Kronrod
is a deterministic algorithm, it does not provide a measure of the
accuracy of the estimation. By contrast, a MC approach
ensures asymptotic convergence for any level of complexity and also produces
precise confidence bands. (See Appendix A for details.)

\subsection{Evaluation and Comparison}
Following the statistical framework used in the original analysis of
this data set \citep{Ginestet2011a}, we tested for the statistical
significance of the $N$-back factor on different topological metrics
using a mixed-effects model. We here have $n=43$ subjects and $J=4$ experimental
conditions. Using the formalism introduced by
\citet{Laird1982}, we have 
\begin{equation}
  \begin{aligned}
    \by_{i}&= \bX_{i}\bbe + \bZ_{i}\Bb_{i} + \bep_{i},
    \qq i=1,\ldots,n, \label{eq:glm1} \\
     &\bep_{i} \stack{\iid}{\sim} N(\bm{0},\sig^{2}_{\ep}\bI) \qq 
     \Bb_{i} \stack{\iid}{\sim} N(\bm{0},\sig^{2}_{b}\bI),
  \end{aligned}
\end{equation}
where $\by_{i}:=[ y_{i1},\ldots,y_{iJ}]^{T}$ is a subject-specific 
vector of topological metrics of interest, $\bbe:=[
\beta_{1},\ldots,\beta_{J}]^{T}$ is a vector of fixed
effect, which do not vary over subjects, $\Bb_{i}:=
b_{i1}$ is a subject-specific random effect and $\bep_{i}:=[
\epsilon_{i1},\ldots,\epsilon_{iJ}]^{T}$ are the residuals. 
Finally, the matrices $\bX_{i}$'s and $\bZ_{i}$'s are given the following
specification, 
\begin{equation}
   \bX_{i} = 
   \begin{bmatrix}
     1 & 0 & 0 & 0 \\
     1 & 1 & 0 & 0 \\
     1 & 0 & 1 & 0 \\
     1 & 0 & 0 & 1 
   \end{bmatrix},\qq\te{and}\qq
  \bZ_{i} = 
   \begin{bmatrix}
     1 \\
     1 \\
     1 \\
     1  
   \end{bmatrix},
\end{equation}
for every $i=1,\ldots,n$. The effect of the $N$-back factor was then
evaluated using Wald's $F$-test. All these analyses were conducted
within the R environment using the lme4 package \citep[see {\rm
  www.cran.r-project.org} and][]{Pinheiro2000}. 
Note that the model used here is slightly simpler than the one used in 
\citet{Ginestet2011a}, as the present mixed-effect model was found to
be better identified than the growth curve model utilized in the
original analysis. 

\begin{figure}[t]
  \includegraphics[width=12cm]{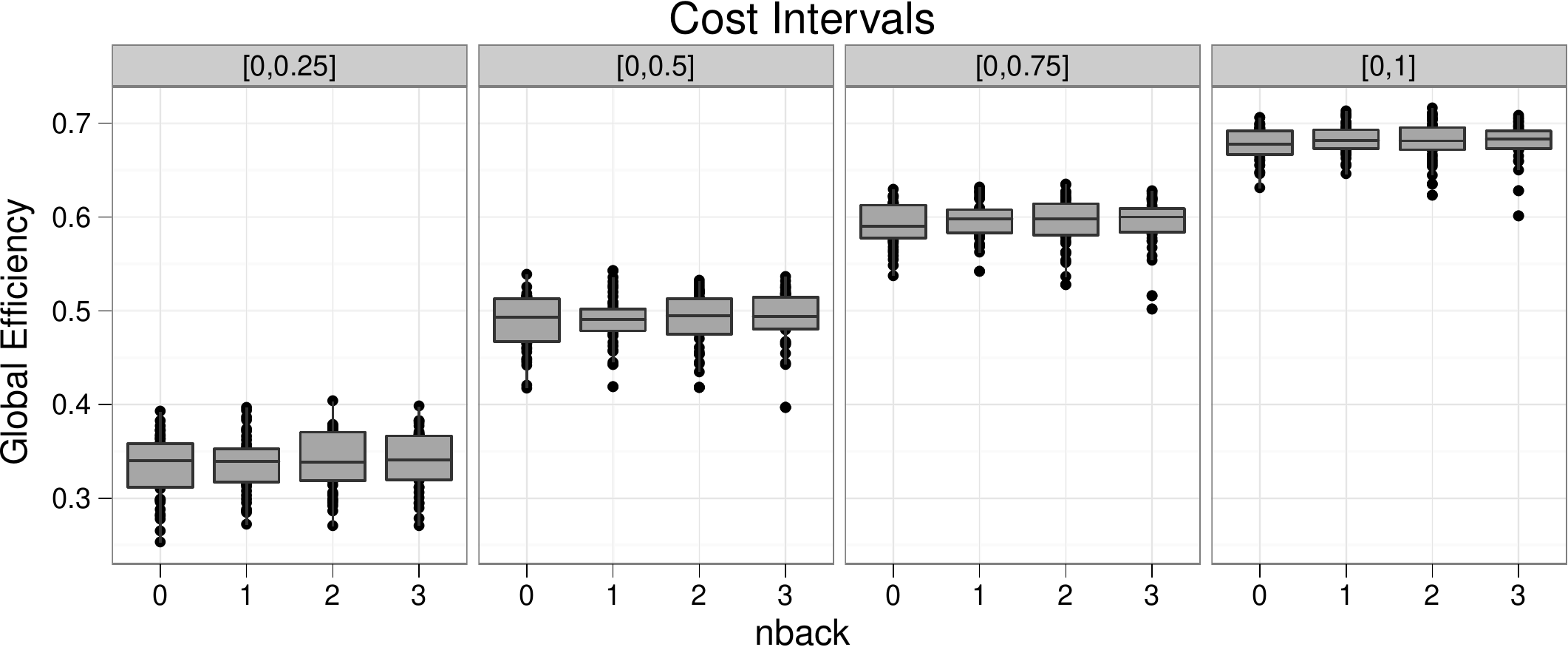}
  \caption{Box plots of cost-integrated global efficiencies 
    of fMRI $N$-back networks for four different
    domains of integration. These integrals were
    estimated using MC approximation over 1,000 samples for
    each of the 43 subjects in each of the four experimental
    condition. Note that different domains of cost-integration do not
    induce any differences in the effect of the experimental
    factor. For all choices of integration domain, there is no
    apparent significant differences. 
   \label{fig:bps}}
\end{figure}

In Figure \ref{fig:bps}, we report the cost--integrated
global efficiencies for this experiment. For illustrative purposes, we
have computed these quantities for four different choices of domains of
integration. The $E_{[0,k]}^{\op{Glo}}(G_{ij})$ were here estimated
using 1,000 MC samples for each subject in each
$N$-back condition. In panel (a), one can observe a clear increase of the
cost-integrated global efficiencies as we increase the size of the
domain of integration, due to the monotonicity of global efficiency
with respect to cost. This is a standard property of global
efficiency: as graphs become denser, their diameter tends to diminish
\citep{Bollobas2003}. In Figure \ref{fig:bps}, one can
also note the dependence of the inter-subject variability of the cost--integrated
metrics on the chosen domain of integration. 
\begin{table}[t]
\begin{center}
\begin{3table}
\small
\caption{Statistical inference for the mixed-effects model described in
  equation (\ref{eq:glm1}) testing for the effect of the $N$-back factor
  on different topological variables. For cost-integrated global
  efficiencies, we have separately tested four different domains of
  integration. \label{tab:tests}}
\begin{tabular}{@{}lllcc@{}}
\toprule
\mcol{2}{c}{Outcome Variable}&\mcol{1}{c}{\ti{Domain}}
&\mcol{1}{c}{$F$-statistic\tnote{a}}&\mcol{1}{c}{$p$-value}\\
\cmidrule(l{5pt}r{5pt}){1-3}\cmidrule(l{5pt}r{5pt}){4-5}\\
\mcol{5}{l}{$K_{W}(G)$}\\
&Weighted Cost & &  3.59  &  0.01  \\
\\
\mcol{5}{l}{$E^{\op{Glo}}_{K}(G)$}\\
&Cost-integrated& $[0,.25]$ &  0.34  &  0.79  \\
&Cost-integrated& $[0,.50]$ &  0.24  &  0.86  \\
&Cost-integrated& $[0,.75]$ &  0.40  &  0.75  \\
&Cost-integrated& $[0,1.0]$ &  1.09  &  0.35  \\
\\
\bottomrule
\end{tabular}
\begin{tablenotes}
        \footnotesize 
        \item[a] Wald $F$-statistic based on model described in
          equation (\ref{eq:glm1}).
\end{tablenotes}
\end{3table}
\end{center}
\end{table}

We therefore tested for the effect of the $N$-back factor on the
topological metrics of interest,
given different domains of integration, in order to evaluate whether
such a choice of domain has a systematic impact on the effect of the
experimental factor. These tests are based on the mixed-effects model
described in equation (\ref{eq:glm1}), and we have reported the results of
these statistical tests in Table \ref{tab:tests}. These results do not
indicate that the choice of different domains of integration yield a
systematic bias in statistical inference. As was reported by 
\citet{Ginestet2011a}, the weighted cost was found to be
systematically affected by the $N$-back factor 
$(\te{Wald }F=3.59,\op{df}_{1}=3,\op{df}_{2}=126,p=0.01)$. However, none of the
cost-integrated global efficiencies appeared to be significantly
influenced by the experimental factor. Most importantly, the use of
different domains of integration did not seem to affect the results.
Integration over the entire cost domain, however, resulted in a larger
$F$-statistic, which may be explained by the lower amount of
variability characterizing cost-integration over larger domains, as
can be observed in Figure (\ref{fig:bps}). 

\begin{figure}[t]
  \includegraphics{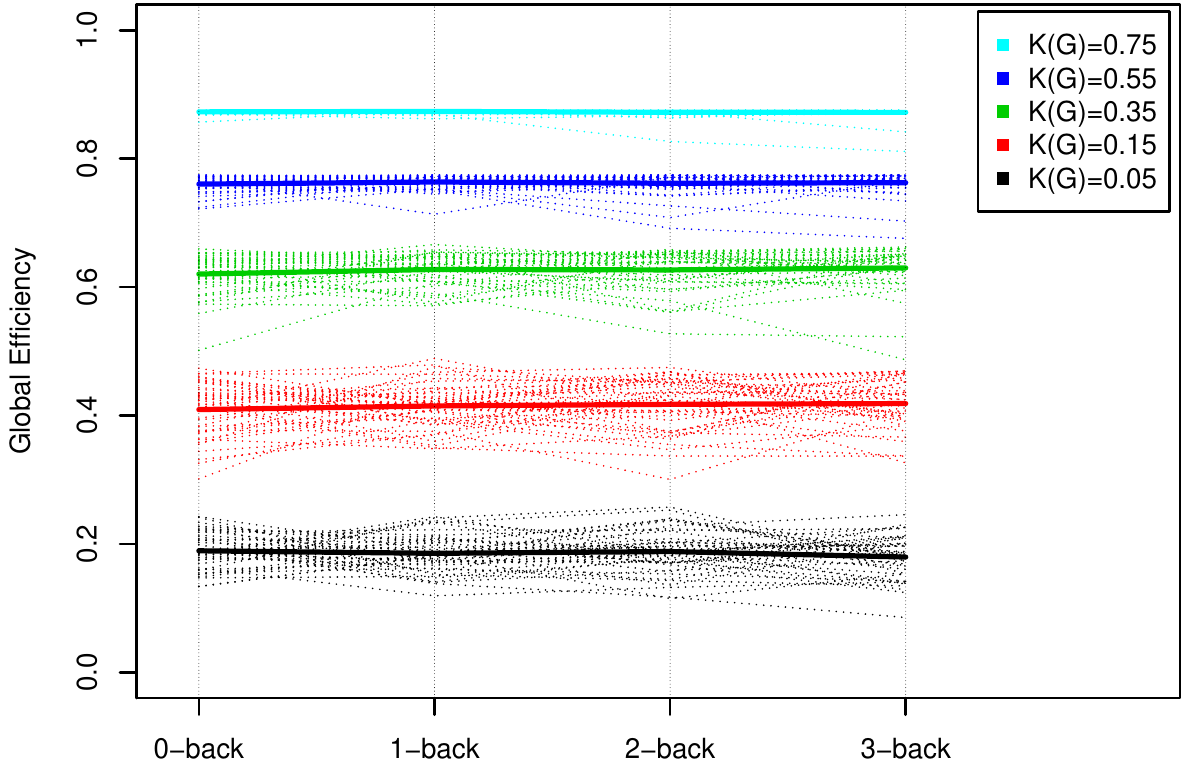}
  \caption{Interaction plots of cost-dependent global efficiencies 
    of fMRI networks with respect to the levels of the $N$-back factor. 
    We here consider five different costs $K\in\lb
    0.05,0.15,0.35,0.55,0.75\rb$. The dashed lines represents the
    cost-specific global efficiencies of each subject, whereas the 
    plain line represents cost-specific global efficiencies averaged over the 43
    subjects. The flatness of the lines at each cost levels suggests
    that the experimental factor has little effect on the topological
    structure of these networks.  
    \label{fig:interaction}}
\end{figure}

In addition, in Table \ref{tab:tests}, we have also reported the
$F$-statistic for the effect of the $N$-back factor on the weighted
cost. 
The subject-specific network's weighted costs were found to be significantly influenced by the
level of the experimental factor, as is immediately visible from the
mean SPNs reported in Figure \ref{fig:net_array}. The separation of
the differences in cost from the differences in topology that results
from the use of a cost-integrated topological metric is best
illustrated by the interaction plots in Figure \ref{fig:interaction}, 
where ensembles of global efficiencies corresponding to different costs
are represented for the four levels of the experimental factors. Note
that, here, we are reporting the efficiency metrics for a single level
of cost, not integrated over a subset of the cost regimen as was done
in Figure \ref{fig:bps}. This is a visual depiction of the $N$-back
factor that corroborates the conclusions reached using cost-integrated
topological metrics, which stated that topology, as measured by global
efficiency, does not significantly vary with the experimental factor. 

\section{Discussion}\label{sec:discussion}
This paper has investigated the effect of thresholding matrices of correlation
coefficients or other measures of association for the purpose of
producing simple unweighted graphs. On the basis of the analysis,
examples and counterexamples studied in
this paper, we make the following methodological recommendations to
researchers intending to compare the topological properties of two or more
populations of weighted networks. 

\subsection{Summary and Recommendations}
We here summarize the main findings of this paper: (i) fixing a
cutoff threshold is not satisfactory, because
this is fully determined by differences in connectivity strength,
as we have shown in section \ref{sec:fixing cutoff}; (ii.) fixing a
subset of cost levels is not satisfactory, because this potentially omits
topological differences at other cost levels (see section
\ref{sec:fixing cost}); (iii) integrating over the entire cost regimen successfully
disentangles connectivity strength from topology up to monotonic
transformations. Specifically, such metrics are invariant to
monotonic transformations of the association weights, as described in 
section \ref{sec:integrating cost}; and (iv.) the weighted topological
metrics, such as $E_{W}$, appear to be too closely related to weighted costs
(see section \ref{sec:weighted metric}). 

From a methodological perspective, we therefore recommend the
following. As a preliminary step, it is good practice to standardize
the association weights,
in order to obtain $w_{ij}\in[0,1]$ for all $w_{ij}$'s, with large
values of the weights corresponding to strong associations. This may
facilitate comparison across separate network analyses, and ease the
interpretation of the results. Secondly, the weighted cost, i.e.~
connectivity strength,
  of the networks of interest can then be computed for all
  networks. This is central to the rest
  of the analysis, and should be conducted systematically. Moreover,
  quantitative differences in connectivity strength \ti{per se} are
  informative about the brain processes at hand, and their
  experimental relevance should not be neglected. 
Thirdly, population differences in cost-integrated topological metrics may then
  be evaluated. This will indicate whether the topologies of the 
  populations under scrutiny vary significantly irrespective of their differences in
  connectivity strength. This aspect of network analysis could be
  regarded as qualitative, as this reflects the networks'
  architectural properties.
We now expand and discuss some of the remarks that were made in
section \ref{sec:integrating cost}. 

\subsection{Limitations of Cost-integration}
As with any form of averaging, cost-integration ignores cost-specific 
topological differences. Networks $G_{1}$ and $G_{2}$, in example
\ref{exa:proportional}, differ in connectivity strength and these differences
may also be expressed through their cost-dependent respective
topologies. That is, as illustrated in example \ref{exa:hybrid},
certain graphs may not exhibit the same topological structure at
different cost levels, and therefore integrating over cost may
potentially mask these subtle topological differences. 
Another potential problem with
cost-integrated quantities is that they may be expensive
computationally. The number of possible cost levels increases at rate
$\cO(N_{V}^{2})$ with respect to the number of vertices in the
networks of interest. In Appendix A and in section \ref{sec:real data},
however, we show how such integrals can be estimated through MC
sampling, which can substantially diminish the required computations.

Another potential pitfall which is not directly visible from proposition
\ref{pro:monotonic} is that the use of cost-integration for the
comparison of several populations of networks requires these networks
to have the same number of positive weights. That is, to be 
comparable two networks do not simply need to possess the same
number of vertices, i.e. $|\cV(G_{1})|=|\cV(G_{2})|$, but also should have
the have the same number of weights, i.e. $|\cW(G_{1})|=|\cW(G_{2})|$.
In this paper, we have re-analyzed an fMRI data set, based on correlation
matrices, which produce fully weighted networks, for which
$N_{I}=N_{W}$ for every subjects. However, when such a condition does
not hold, we recommend the selection of a domain of integration that
corresponds to the smallest common denominator. That is, 
$N_{W}\as:= \min_{i=1,\ldots,n} |\cW(G_{i})|$, for a given population
of $n$ weighted networks denoted $G_{i}$. Thus, when considering
sparser networks, such as structural brain networks, one may still be
able to control for differences in cost, by integrating over a subset
of the cost regimen, which reflects the sparsity of the networks under
comparison. 

A similar problem may arise if one or several networks in the
population of interest have multiplicities, i.e. weights that take
identical values. Since 
cost-integration relies on the ranking of weights, it follows that one
may need to adjust for such multiplicities, otherwise this can lead
to spurious generation of random topologies. That is, when the tied
ranks are resolved by random ordering, the allocation
of weights with identical values to specific cost levels is random,
and therefore artificially create a random topology for these cost
levels. For sparse networks, multiplicities are likely to
arise around zero. However, for large non-sparse networks, the
occurrence of multiplicities should be evaluated by counting the
number of tied ranks in the distribution of the weights. In
particular, if the two populations of networks that one wishes to
compare differ significantly in number of tied ranks, then
comparison based on cost-integration will be contaminated by an
artificial level of random topology.

Another possible limitation of cost-integration is that by integrating
over several cost levels, we omit to take into account the dependence
between the topologies of the different thresholded graphs. The
topological structure of the unweighted networks created by
thresholding the original weighted graph share the same
edges. Arguably, the cumulative nature of this procedure results in
emphasizing the importance of the set of edges with the largest
weights. Once these edges have been included into a thresholded graph, 
they will be retained for the remaining cost levels. This is
especially true for the topological metrics that we have studied in
this paper, since global and local efficiencies are both monotonic
functions of cost. 

Finally, one may also be interested in addressing how
differences in cost and differences in topology interact. By
controlling for monotonic differences in wiring cost, we potentially
ignore how cost differences may contribute to the topological
structure. In the supplementary methods, we report a different
type of integrated topological metrics, which attempts to combine cost
and topological differences. In this case, cost-integration was
weighted with respect to the distance between each pair of weights used
for thresholding the graph (see the supplementary
methods document). Unfortunately, we have shown that this choice of integration exhibits
some undesirable properties, in the sense that it tends to give more
importance to very low- and very high-cost topologies. Further
research will therefore be required to produce more relevant topological
functions of weighted graphs, which provide a better understanding of
the interaction between cost and topology. 

\subsection{Future Work}
Most of this paper has focused on the global efficiency metric. 
Thus, our conclusions and the examples studied will not necessarily
apply to other topological measures. However, our main result (proposition
\ref{pro:monotonic}) was proved in a very general setting, which is
independent of the particular formula of the topological metric of
interest. Our general conclusion about the usefulness of
cost-integration when one wishes to disentangle differences in cost
from differences in topology is therefore valid for any topological
metric defined for an unweighted graph. 
In addition, we note that since most weighted metrics are constructed
on the basis of the  weighted shortest path matrix, one may surmise
that our second main theoretical 
result (proposition \ref{pro:weighted-cost}), may hold in a more
general setting. However, a proof that the  equivalence relationship
between the weighted version of a topological metric and 
the weighted cost, for instance, hold for topological measures other
than the global efficiency would require further work. 

Thus far, we have only considered flat distributions on the space of
network costs. However, future methodological developments will be needed in order to consider
more sophisticated approaches to this problem. In particular,
the specification of a prior distribution on $K$ should take 
into account the effect size associated with different values of this
random variable. When considering correlation coefficients, for
instance, it can easily be shown that higher values indicate larger
effects, and it may therefore
be preferable to emphasize network comparisons built upon the largest
correlation coefficients. This may be implemented by integrating
network topological metrics with respect to a skewed distribution on
$K$, which puts more weight on sparse networks, whose edges are better
identified. 

One should note that the use of cost-integration when comparing
weighted networks is not akin to taking into consideration the
multilevel or hierarchical nature of a weighted network. Such a structural interpretation
of the successive thresholding necessary for such an integration is not
necessary to justify the usefulness of the method in controlling for
monotonic differences in weighted cost. Since the networks of interest `exist'
as weighted networks, their thresholding remains artificial and it is
not clear whether one can ascribe any substantive meaning to the
resulting family of thresholded graphs. Further work will therefore 
be needed in order to better characterize the architecture of the 
ensemble of thresholded discrete networks subtending a weighted graph.

\section{Acknowledgments} 
This work was supported by a fellowship from the UK National Institute
for Health Research (NIHR) Biomedical Research Centre for Mental
Health (BRC-MH) at the South London and Maudsley NHS Foundation Trust
and King's College London. This work has also been funded by
the Guy's and St Thomas' Charitable Foundation as well as the South
London and Maudsley Trustees. We would also like to thank 
three anonymous reviewers for their valuable inputs in improving this
manuscript. 

\pagebreak
\setstretch{1} 
\small
\setcounter{secnumdepth}{-2}
\section{Appendices}
\subsection{A: Monte Carlo (MC) Sampling}
The cost-integrated quantities introduced in section
\ref{sec:cost-integrated} may first appear unwieldy to compute, especially
when considering large graphs. However, the structure of these integrals allows the
construction of a straightforward MC sampling scheme. 
This classical approximation method has the advantage of providing
both an estimate of the quantity
of interest and an estimate of the variance of that estimation. For an
introductory text to MC techniques, see
\citet{Gilks1996}, and for a more advanced treatment,
\citet{Robert2004}. 

In order to apply MC sampling theory, we first observe that
our integration problem --that is, the computation of $E_{K}$-- can
be re-formulated as an expectation. For convenience, we drop any
reference to the function $\gamma(G,K)$, and therefore denote
the efficiency metric $E(\gamma(G,K))$ as $E(K)$. The cost-integrated
metric $E_{K}$ can then be expressed as an expectation of $E(K)$ since,
straightforwardly, we have
\begin{equation}
      E_{K} = \int_{\Omega_{K}}E(k)p(k)dk = \E_{p}\lt[E(K)\rt],
      \label{eq:cost-integral mc}
\end{equation}
where $\Omega_{K}$ is the space of all possible costs for $G$, with 
$|\Omega_{K}|=\binom{N_{V}}{2}=N_{V}(N_{V}-1)/2$.
The expectation in (\ref{eq:cost-integral mc}) is taken with respect
to $p$, the probability density function of $K$, and explicit reference to $G$ has been
omitted. It is natural to consider the use of a sample
$\lb k_{1},\ldots,k_{m}\rb$ from $p$ in order to approximate 
$E_{K}$ by the following empirical average, 
\begin{equation}
      \overline{E}^{(m)}_{K} = \frac{1}{m}\sum_{l=1}^{m} E(k_{l}).
\end{equation}
The approximation $\overline{E}^{(m)}_{K}$ converges to $E_{K}$
almost surely, by the Strong Law of Large Numbers. In addition,
providing that $E(K)$ is
square-integrable, the speed of convergence of
$\overline{E}^{(m)}_{K}$ can be evaluated by considering the
theoretical variance of that estimate, 
\begin{equation}
     \var\lt( \overline{E}^{(m)}_{K}\rt) = \frac{1}{m}\int\limits_{[0,1]}
          \Big(E(k)-\E_{p}\lt[E(k)\rt]\Big)^{2}p(k)dk.
\end{equation}
which can be approximated by the following MC variance, 
\begin{equation}
     \lt(\sig^{(m)}_{K}\rt)^{2} = \frac{1}{m^{2}}\sum_{l=1}^{m} 
         \lt(E(k_{l})- \overline{E}^{(m)}_{K}\rt)^{2}.
\end{equation}
This quantity is of special interest in MC sampling, as it
permits the evaluation of the rate of convergence of the estimation. 
It is generally referred to as the \ti{MC standard error}.
Using Slutsky's theorem, it can also be shown that as $m\to\infty$,
the random variable, 
\begin{equation}
    \frac{\overline{E}^{(m)}_{K} - E_{K}}{\sig_{K}^{(m)}},
\end{equation}
has the probability density function of a Normal variate centered at
zero, with unit variance. MC sampling is especially useful when the
stochastic function that we wish to integrate --here, denoted $E(K)$--
is complex, whereas the random variable with
respect to which we integrate can easily be sampled. Most
topological metrics will be of a complex nature --i.e.~ non-linear. By
contrast, both $K$ and $C$ will be straightforward to
sample, since we have specified uniform distributions for both of them.
The theory underlying MC sampling is general and
can therefore be applied to any type of topological metrics. Care,
however, should be taken when evaluating the properties of very large
networks, where the topology may vary substantially from one level of
cost to another. When confronted with such large networks, however,
the MC standard error is still a good indicator of
the accuracy of such approximations. 

\subsection{B: Proof of Proposition \ref{pro:monotonic}}
In order to prove proposition (\ref{pro:monotonic}), we first need to
give a formal definition of $\gamma(G,k)$, for some given weighted
network $G=(\cV,\cE,\cW)$. This function relies on the concept of
rank, which can be formally defined in our context, as follows
\begin{equation}
     R_{ij}(\bW):= \frac{1}{2}\sum_{u=1}^{N_{V}}\sum_{v\neq u}^{N_{V}}
     \cI\lb w_{ij}\leq w_{uv}\rb,
\end{equation}
where $R_{ij}=1$ implies that $w_{ij}$ is the largest weight in $\bW$.
Here, we have assumed that there are no ties in the ranks of $\bW$. When
ties occur in practice, we recommend to resolve tied ranks by
assigning the corresponding ordering of the elements' indices. By
contrast, resolving tied ranks using random allocation can result in
introducing a spurious amount of random topology in the networks of
interest. The presence of tied ranks, however, will generally be
indicative of a high level of sparsity, which is better dealt with by
restricting the domain of integration. 

Computationally, this definition can be simplified if one only considers
the upper off-diagonal elements of $\bW$ and omits the division by 2. For our
purpose, this definition will be more convenient. These ranks can be
standardized in order to derive the \ti{percentile ranks}, 
\begin{equation}
    P_{ij}(\bW) := \frac{R_{ij}(\bW)}{N_{I}}, 
\end{equation}
where $N_{I}$ is the number of edges in the saturated version of
$G$. Note that the resulting matrices $\bR$ and $\bP$ of ranks and
percentile ranks, respectively, are both symmetric. A good introduction to order
statistics, ranks and percentile ranks is provided by \citet{Lin2006}.

The function $\gamma(G,k)$ can now be given a formal definition using
the $P_{ij}$'s, such that
\begin{equation}
      \gamma(G,k) :=: \gamma(\bW,k) := \cI\lb \bP(\bW) \leq k\rb,
      \label{eq:gamma}
\end{equation}
where the indicator function is applied elementwise to matrix
$\bP(\bW)$, where $\bW$ is the similarity matrix of $G$. 
It can hence be seen that the function $\gamma$ prescribes an
adjacency matrix $\bA^{(k)}$ with the desired cost. This can be verified by
computing the cost of the corresponding unweighted network
$G^{(k)}=(\cV,\cE^{(k)})$, where $\cE^{(k)}$ is the edge set that
populates $\bA^{(k)}$, obtained after 
application of the $\gamma$ function at $k$. Provided that $k\in \Omega_{K}$, as
defined in equation (\ref{eq:costs}), we have 
\begin{equation}
     K(G^{(k)}) = \frac{1}{N_{I}} \sum_{\cI(G)} a^{(k)}_{ij} 
            = \frac{1}{N_{I}} \sum_{\cI(G)}\cI\lb P_{ij}(\bA^{(k)})\leq k\rb = k,     
     \label{eq:K-to-k}
\end{equation}
which can be verified by noting that equation (\ref{eq:K-to-k}) is
simply the discrete version of the integration of an indicator
function of the form, $\int_{0}^{1}\cI\lb x\leq k\rb
dx=\int_{0}^{k}dx=k$. Using this notation, the proof of proposition
\ref{pro:monotonic} is now straightforward. This demonstration uses the
fact that a monotonic function does not modify the ranks of its arguments. 
\begin{proof}
    Recall that the cost-integrated version of $E(G)$, in its
    computational form, is given by
    \begin{equation}
          E_{K}(\bW)=\frac{1}{N_{I}}\sum_{t=1}^{N_{I}-1}E(\ga(\bW,k_{t})). 
    \end{equation}
     To demonstrate that $E_{K}(\bW) = E_{K}(h(\bW))$, it therefore
     suffices to show that 
     \begin{equation}
          E_{K}(\gamma(\bW,k_{t})) = E_{K}(\gamma(h(\bW),k_{t})),
     \end{equation}
     for every $k_{t}$, which further simplifies to the sole requirement
     that $\gamma(\bW,k_{t})=\gamma(h(\bW),k_{t})$, for all
     $t=1,\ldots,N_{I}$. From the definition of the $\gamma$ function
     introduced in equation (\ref{eq:gamma}), we have the following relationship,
     \begin{equation}
          \gamma(h(\bW),k_{t}) = \cI\lb \bP(h(\bW)) \leq k\rb = 
                \cI\lt\lb  \frac{R_{ij}(h(\bW))}{N_{I}} \leq k \rt\rb.
     \end{equation}
     However, one can observe that, since $h$ is applied elementwise, we have 
     \begin{equation}
          R_{ij}(h(\bW)) = \frac{1}{2}\sum_{u=1}^{N_{V}}\sum_{v\neq u}^{N_{V}} \cI\lb
                 h(w_{ij})\leq h(w_{uv})\rb = R_{ij}(\bW),
          \label{eq:rank proof}
     \end{equation}
     for any monotonic function $h$. Note that this argument makes no
     use of the definition of $E$. This completes the proof. 
\end{proof}

\subsection{C: Proof of Proposition \ref{pro:weighted-cost}}
\begin{proof}
   We prove the result by contradiction. Assume that the conclusion
   does not hold. That is, $E_{W}\neq K_{W}$. 
   By applying the definitions of $E_{W}$ and $K_{W}$ in
   equations (\ref{eq:weighted definition}) and (\ref{eq:weighted cost}), respectively, we have
   \begin{equation}
         E_{W}(G) := \frac{1}{N_{I}}\sum_{\cI(G)} \frac{1}{d^{W}_{ij}} \neq 
              \frac{1}{N_{I}}\sum_{\cI(G)}w_{ij} =: K_{W}(G).
   \end{equation}
   It therefore suffices to show that $d^{W}_{ij}\neq w^{-1}_{ij}$ for at
   least one of the weights. The weighted shortest path $d^{W}_{ij}$
   is defined in equation (\ref{eq:weighted shortest path}) as 
   \begin{equation}
       d^{W}_{ij} := \min_{P_{ij}\in\cP_{ij}(G)}\sum_{w_{uv}\in
                    \cE(P_{ij})} w^{-1}_{uv}.
   \end{equation}
   It follows that $d^{W}_{ij}\neq w^{-1}_{ij}$ if and only if there
   exists a path $P^{\as}_{ij}$ in $\cP_{ij}(G)$, which satisfies 
   \begin{equation}
        \sum_{w_{uv}\in\cE(P^{\as}_{ij})} w^{-1}_{uv} < w_{ij}^{-1}.
        \label{eq:proof E1}
   \end{equation}
   That is, the path $P^{\as}_{ij}$ is shorter than the direct
   connection $w_{ij}$ between the $i\tth$ and $j\tth$
   vertices. Inequality (\ref{eq:proof E1}) can be sandwiched in the following
   fashion, 
   \begin{equation}
        |\cE(P^{\as}_{ij})|\lt(\max_{w_{ij}\in\cE(G)} w_{ij}\rt)^{-1}\leq
        \sum_{w_{uv}\in\cE(P^{\as}_{ij})} w^{-1}_{uv} < w_{ij}^{-1} \leq
        \lt(\min_{w_{ij}\in\cE(G)} w_{ij}\rt)^{-1},
   \end{equation}
   where $|\cE(P^{\as}_{ij})|$ denotes the cardinality of $P^{\as}_{ij}$.
   Inverting the entire inequality then gives
   \begin{equation}
        \frac{1}{|\cE(P^{\as}_{ij})|}\max_{w_{ij}\in\cE(G)} w_{ij} \geq
        \lt(\sum_{w_{uv}\in\cE(P^{\as}_{ij})} w^{-1}_{uv}\rt)^{-1} > w_{ij} \geq
        \min_{w_{ij}\in\cE(G)} w_{ij}.
   \end{equation}
   However, we clearly have
   \begin{equation}
        \frac{1}{2}\max_{w_{ij}\in\cE(G)} w_{ij} \geq
        \frac{1}{|\cE(P^{\as}_{ij})|}\max_{w_{ij}\in\cE(G)} w_{ij} > \min_{w_{ij}\in\cE(G)} w_{ij},
   \end{equation}
   which contradicts our hypothesis, and proves the claim.
\end{proof}

\small
\singlespacing
\addcontentsline{toc}{section}{References}
\bibliography{/home/cgineste/ref/bibtex/Statistics,%
             /home/cgineste/ref/bibtex/Neuroscience}
\bibliographystyle{oupced3}


\addcontentsline{toc}{section}{Index}

\end{document}